\documentclass[useAMS,fleqn,usenatbib]{mnras}
\usepackage{amssymb}
\usepackage{graphicx}

\usepackage{tikz}

\tikzset{cross/.style={cross out, draw, 
		minimum size=2*(#1-\pgflinewidth), 
		inner sep=0pt, outer sep=0pt}}

\makeatletter
\define@key{Gin}{rotate}{}
\makeatother

\newcommand{\nr}[1]{{#1}}

\newcommand{\lp}{\ensuremath{\left(}}
\newcommand{\rp}{\ensuremath{\right)}}
\newcommand{\lb}{\ensuremath{\left[}}
\newcommand{\rb}{\ensuremath{\right]}}

\newcommand{\abs}[1]{\ensuremath{\left| #1 \right|}}

\newcommand{\at}[2]{\ensuremath{\left. #1 \right|_{#2}}}

\renewcommand{\xi}{\ensuremath{\bmath{x}_{i}}}
\newcommand{\xj}{\ensuremath{\bmath{x}_{j}}}
\newcommand{\xij}{\ensuremath{\bmath{x}_{ij}}}
\newcommand{\Wij}{\ensuremath{W_{ij}}}

\newcommand{\tr}[1]{\ensuremath{\mbox{tr} \lp #1 \rp}}
\newcommand{\unity}{\ensuremath{\bmath{1}}}

\newcommand{\Aab}{\ensuremath{A_{\alpha \beta}}}

\newcommand{\ngbsum}{\ensuremath{\sum\limits_{j=1}^{N_{ngb}}}}

\newcommand{\infint}{\int\limits_{-\infty}^{\infty}}

\newcommand{\erf}{\mbox{erf}}

\title{Anisotropic thermal conduction in galaxy clusters with MHD in Gadget}

\author[A. Arth et al.]
{A.~Arth$^{1,2}$\thanks{E-mail: arth@usm.uni-muenchen.de}, K. Dolag$^{1,3}$, A. M. Beck$^{1}$, M.~Petkova$^{1,4}$ and H.~Lesch$^1$\\
$^{1}$University Observatory Munich, Scheinerstr. 1, D-81679 Munich, Germany\\
$^{2}$Max Planck Institute for Extraterrestrial Physics, Giessenbachstr. 1, D-85748 Garching, Germany\\
$^{3}$Max Planck Institute for Astrophysics, Karl-Schwarzschild-Str. 1, D-85741 Garching, Germany\\
$^{4}$Excellence Cluster Universe, Boltzmannstr. 2, D-85748 Garching, Germany}
	
\date{Accepted ???. Received ???; in original form ???}

\begin{document}
\pagerange{\pageref{firstpage}--\pageref{lastpage}} \pubyear{2014}

\maketitle

\label{firstpage}

\begin{abstract}
	We present an implementation of physically motivated thermal conduction including the anisotropic effects of magnetic fields for smoothed particle hydrodynamics (SPH).
	The diffusion of charged particles and therefore thermal conduction is mainly proceeding parallel to magnetic field lines and suppressed perpendicular to them.
	We derive an SPH formalism for the anisotropic heat transport and solve the corresponding equations with an implicit conjugate gradient scheme.
	We discuss several issues of unphysical heat transport in the cases of extreme ansiotropies or unmagnetised regions and present possible numerical workarounds.
	
	We implement our algorithm into the GADGET code and study its behaviour in several 3D test cases.
	In general, we reproduce the analytical solutions of our idealised test problems, and obtain good results in cosmological simulations of galaxy cluster formations.
	Within galaxy clusters, the anisotropic conduction produces a net heat transport similar to an isotropic Spitzer conduction model with an efficiency of one per cent.
	Compared to observations, isotropic conduction with more than 10 per cent of the Spitzer value leads to an oversmoothed temperature distribution within clusters, while the results obtained with anisotropic conduction reproduce observed temperature fluctuations well.
	
	A proper treatment of heat transport is crucial especially in the outskirts of clusters and in high density regions.
	It's connection to the local dynamical state of the cluster also might contribute to the observed bimodal distribution of (non) cool core clusters.
	Our new scheme significantly advances the modelling of thermal conduction in numerical simulations and overall gives better results compared to observations.
\end{abstract}


\section{Introduction}
\label{introduction}
	Galaxy clusters are the largest gravitationally bound systems in our Universe.
	They typically contain hundreds of galaxies and are detected in a broad range of frequencies.
	In particular, we observe strong X-ray radiation from the hot intra cluster medium (ICM), which is mainly caused by continuous free-free emission of thermal electrons and discrete metal lines.
	Observations show that clusters contain roughly ten per cent of their total mass in a hot baryonic gas component with temperatures up to several $10^8$ K (corresponding to about 15 keV) at typical densities of $10^{-3}$ cm$^{-3}$.
	
	The state of the ICM and its related physical properties are crucial to investigate and learn about the formation and evolution of galaxy clusters.
	The gas is partially heated at the virial shock by the gravitational infall and furthermore raised in temperature by shocks within the ICM.
	Galactic winds driven by the star formation process and feedback of active galactic nuclei (AGN) inject additional kinetic and thermal energy to the hot cluster medium.
	Radiative processes within the ICM, such as thermal bremsstrahlung or metal lines \citep{Peterson2006} allow the gas to cool and loose energy.
	Furthermore, all galaxy cluster are known to host magnetic fields with strengths up to several $\mu$G \citep{Kronberg1994,Taylor2006}, which influences the dynamics of the hot plasma.
	The structure and the amplitude of the cluster magnetic fields guide the propagation of charged particles and contribute to the equation of motion via the Lorentz force.
	The magnetic field lines are assumed to be highly tangled and twisted on small scales, which then will contribute of the heating of the plasma by magnetic reconnection \citep[e.g.][]{Fabian1994}.
	Little is know about the magnetic field structure as its mapping is observationally challenging \citep{Fabian2002,Sarazin2008,Komarov2013}.
	
	We discuss the influence of the anisotropic magnetic field on plasma transport processes and in particular on thermal conduction.
	From a microscopic point of view thermal conduction is the heat transport due to collisions of electrons.
	Therefore, it strongly depends on temperature, which allows us to use probes of the hot ICM to study large-scale transport of heat.
	From a macroscopic point of view thermal conduction can be modelled as a diffusion process re-distributing internal energy.
	Usually, the so called Spitzer conduction \cite[see][]{Spitzer1956} is used as a general formulation and multiplied with an efficiency factor, which changes with and depends on the astrophysical environment.
	This original formulation assumes an isotropic movement of electrons and corresponding distribution of collisions.
	As galaxy clusters host significantly strong magnetic fields the influence of the magnetic fields onto thermal conduction has to be taken into account.
	Since particle movement perpendicular to magnetic field lines is restricted, the assumption of isotropic collisions does not hold any longer and thermal conduction is coupled to the magnetic field topology.
	This result in different heat transport timescales parallel and perpendicular to the magnetic field.
	However, the magnetic fields needs to be sufficiently strong to dominate the mean free path \citep{Rechester1978}.
	
	Observational evidence in galaxy clusters of suppressed perpendicular heat transport is found in so called cold fronts, i.e. regions with a rather stable temperature gradient, but no pressure gradient \citep{Owers2009,Owers2011}.
	This phenomenon demonstrates the insulation of gas with respect to conduction most probably through magnetic fields \citep{Peterson2003}.
	Moreover, turbulent magnetic fields also become a very interesting case to study giving rise to different ways to estimate efficiency factors averaging the suppression effect over volume.
	For example \cite{Chandran1998} proposed that Spitzer conduction should be suppressed by a factor of about 1/300, while \cite{Narayan2001} claimed that conduction can still be efficient up to 1/5 of the Spitzer value in highly turbulent environments.
	Recent studies of density and velocity power spectra also suggest a significant amount of conductivity in galaxy clusters. \citep{Gaspari2014}
	
	Thermal conduction was frequently discussed as a heating source to balance cooling losses in galaxy clusters in order to explain why hot gas is observed despite the cooling times being smaller than the cluster life times.
	However, the impact of this effect is questioned and even less significant, when the anisotropies of the magnetic field are included.
	For detailed discussions of this problem we refer to \cite{Binney1981}, \cite{Bregman1989}, \cite{Fabian2002}, \cite{Loeb2002}, \cite{Zakamska2003}, \cite{Voigt2004}, \cite{Fabian2006} and \cite{Rasia2015}.
	
	In pioneering work, usually a factor of 1/3 is multiplied onto the Spitzer value as a correction for the influence magnetic fields (e.g. \citealt{Dolag2004}, supported by work presented in \citealt{Rosner1989}).
	We present a numerical implementation of the anisotropic heat transport equation including the seeding and evolution of magnetic fields applied to cosmological simulations of galaxy clusters.
	
	A solver for anisotropic thermal conduction is already part of several commonly used grid based codes, which solve the equations of magnetohydrodynamics (MHD).
	For implementation details and cluster simulations see for example \cite{Parrish2005}, \cite{Bogdanovic2009}, \cite{Avara2013} (ATHENA code) or \cite{Ruszkowski2011} (FLASH code).
	\nr{Recently \cite{Hopkins2016} presented the implementation of diffusion equations for the hybrid code GIZMO.}
	However, we present the first implementation of anisotropic thermal conduction into a smoothed particle magnetohydrodynamics (SPMHD) code.
	In contrast to the Eulerian methods, which discretise the volume, SPH discretises the mass and is commonly used in simulations of structure formation.
	The Lagrangian nature of SPH ensures the conservation of energy, momentum and angular momentum and allows to resolve large density gradients.
	For a recent review on the SPH method we refer to \cite{Price2012}.
	
	In this paper we present simulations performed with the N-body / SPH code GADGET \citep{Springel2001,Springel2005a} with the non-ideal MHD implementation based on \cite{Dolag2009}, \cite{Stasyszyn2013} and \cite{Bonafede2011}.
	The SPH code evolves entropy as the thermodynamical variable of choice \citep{Springel2002}.
	Additionally, we include several improvements for SPH such as the Wendland kernel functions \citep{Dehnen2012}.
	More information on some of the SPH improvements can be found in \citet{Beck2015}.
	Our cosmological simulations include sub-grid models for radiative cooling, star formation and supernova feedback as described by \cite{Springel2003}. 
	Furthermore, we employ a supernova seeding scheme for the magnetic field \citep{Beck2013} in contrast to previous simulations using uniform initial magnetic fields \citep[e.g.][]{Dolag1999,Beck2012}.
	Therefore, we start without any initial magnetic field in all of the simulations.
	Magnetic fields are then frequently seeded (i.e. injected) during the simulation within star-forming regions by supernova events and then further evolved by the MHD equations \citep{Dolag2009}.
	We use and extend the existing implementation of isotropic thermal conduction by \cite{Jubelgas2004} with the conjugate gradient solver described in \cite{Petkova2009}.
	
	The paper is structured as follows.
	In section \ref{phenomenologyconduction} we explain the physics behind (anisotropic) thermal conduction, and describe our SPH implementation in section \ref{numericalimplementation}.
	Section \ref{tests} presents several test problems before we analyse simulations of galaxy cluster formation in section \ref{applicationgeneral}.

\section{Phenomenology of thermal conduction}
	\label{phenomenologyconduction}

	We start with a brief introduction in the physical properties and concepts of isotropic as well as anisotropic conduction.
	
	\subsection{Review of isotropic thermal conduction}
	
		According to \cite{Spitzer1956} we can write down a conduction heat flux resulting from a temperature gradient using Fourier's law as
		\begin{equation}
			\bmath{Q} = - \kappa \bmath{\nabla} T
			\label{EQbasicheatflux}
		\end{equation}
		with the conduction coefficient $\kappa$. For an idealized Lorentz gas we can assume Spitzer conductivity, which equals a coefficient of
		\begin{equation}
			\kappa_{\mathrm{Sp}} = 20 \lp \frac{2}{\pi} \rp^{3/2} \frac{\lp k_B T_e \rp ^{5/2} k_B}{m_e^{1/2} e^4 Z ~ \ln \Lambda},
		\end{equation}
		with $Z$ the average proton number of the plasma, the Coulomb logarithm $\ln \Lambda$ and electron temperature $T_e$, mass $m_e$ and elementary charge $e$. Most important is the strong dependence of conductivity on the electron temperature. Therefore, we assume that thermal conduction has an important influence mainly on very hot gas, as for example in the central regions of massive galaxy clusters where the plasma reaches temperatures up to about $10^8 \mathrm{K}$.
		
		In fact, we need to multiply the idealized Spitzer conductivity by an additional factor $\delta$:
		\begin{equation}
			\kappa = \delta \cdot \kappa_{\mathrm{Sp}}.
		\end{equation}
		This factor has been calculated by \cite{Spitzer1953} and is highly dependent on the average proton number of the plasma. $\delta = 0.225$ for a pure proton electron plasma and rises up close to 1 for large values of $Z$.
		
		\cite{Spitzer1953} describe a way to calculate the average proton number by summation over all ions in the plasma. Applying a primordial hydrogen - helium plasma, they find a value of $Z \approx 1.136$. Using the tabulated values in \cite{Spitzer1956} we obtain a factor of $\delta \approx 0.3$.
		
		When used for cosmological simulations one often assumes a primordial gas distribution. A typical value for an effective conductivity $\kappa$ is e.g. given by \cite{Sarazin1986}
		\begin{equation}
			\kappa = 1.31 \cdot n_e \lambda_e k_B \lp \frac{k_B T_e}{m_e} \rp^{1/2},
		\end{equation}
		or
		\begin{equation}
			\kappa = 4.6 \cdot 10^{13} \lp \frac{T_e}{10^8 K} \rp^{5/2} \frac{40}{\ln \Lambda} ~~~ \frac{\mathrm{erg}}{\mathrm{s ~ cm ~ K}}
		\end{equation}
		with the electron number density $n_e$ and the mean free path of the electrons $\lambda_e$. Because of the inverse dependence on mass we infer that electrons give a much stronger contribution to heat conduction than protons. This is reasonable since lighter particles have higher thermal velocities at a fixed temperature and can be accelerated much easier. Therefore, the amount of collisions in a given time span are drastically increased for low mass particles. Consequently, only electrons are considered in the calculation and we omit the index $e$ in our equations hereafter.

		We neglect any dependency of the Coulomb logarithm on temperature and electron density and use $\ln \Lambda = 37.8$, which is a fairly good approximation for typical plasmas in our study. More precise calculations for different collision events (e.g. electron-electron or electron-proton) can be found for example in \cite{Huba2011}. What remains is the strong dependence on temperature to the power of $5/2$.
		
		Furthermore, we need to apply an important correction. So far we assumed, that the typical length scale of the temperature gradient $l_T = T/\abs{\bmath{\nabla} T}$ is always much larger than the mean free path. However, for very low density plasmas one cannot expect a high conductivity even if the temperature rises a lot, since scatterings and therefore energy transfer events happen only at a very low rate. \cite{CowieLennoxL.McKee1977} have calculated the saturated heat flux for this case as
		\begin{equation}
			Q_{\mathrm{sat}} = 0.4 n_e k_B T \lp \frac{2 k_B T}{\pi m} \rp^{1/2}.
			\label{EQsaturatedflux}
		\end{equation}
		Interpolating between these findings and the common Spitzer conduction coefficient we estimate a corrected heat flux as
		\begin{equation}
			Q_ {\mathrm{tot}} = - \frac{\kappa \cdot T}{l_T + 4.2 \lambda} \frac{\bmath{\nabla} T}{\abs{\bmath{\nabla} T}}.
		\end{equation}
		Alternatively, we can redefine the conduction coefficient as
		\begin{equation}
			\kappa = \frac{\kappa_{\mathrm{Sp}}}{1 + 4.2 \lambda / l_T}.
		\end{equation}
		This modified Spitzer conduction is applicable to galaxy clusters and giant elliptical galaxies if magnetic fields are not taken into account. For a detailed discussion we refer to \cite{Rosner1989}.
		
		Finally, we can write the effect of thermal conduction as a change of specific internal energy
		\begin{equation}
			\frac{du}{dt} = - \frac{1}{\rho} \bmath{\nabla} \cdot \bmath{Q} = \frac{1}{\rho} \bmath{\nabla} \cdot \lp \kappa \bmath{\nabla} T \rp ,
		\end{equation}
		which depends on the density $\rho$ and on the heat flux directed anti-parallel to the temperature gradient.
	
	\subsection{Description of anisotropic conduction}
		\label{anisotropicthermalconduction}
		
		Next, we add magnetic fields to the picture. As previously mentioned, thermal conduction is based on Coulomb collisions of charged particles. Except these collisions particles are allowed to move freely in the plasma. However, in the presence of magnetic fields the movement perpendicular to the field lines is restricted. The electrons move on spiral trajectories around the field lines. The frequency of the circular motion, which depends on the strength of the magnetic field $B$, is called Larmor- or gyrofrequency:
		\begin{equation}
			\omega_g = \frac{e B}{m c}.
		\end{equation}
		To see how this affects the capability of electrons to transport energy, we present some phenomenological ideas and scaling relations on how a general electron diffusion process is affected by magnetic fields, following \cite{Frank-Kamenezki1967}. A detailed derivation and discussion of the plasma physics behind this simplistic approach can be found for example in \cite{Braginskii1965}, who presents three terms of conductive heat flux:
		\begin{equation}
			Q = - \kappa_\parallel \nabla_\parallel T - \kappa_\perp \nabla_\perp T - \kappa_\Lambda \hat{B} \times \nabla T
			\label{EQconductionbraginskii}
		\end{equation}	
		\nr{We investigate how the perpendicular term $\kappa_\perp$ scales with the magnetic field strength in the following two subsections and shortly talk about the hall term $\kappa_\Lambda$ afterwards in section \ref{physicalhall}.}
		
		Due to the similar microscopic origin we can infer the following relations for diffusion to hold also for thermal conduction. This connection is motivated through some scaling relations starting with the ideal gas law
		\begin{equation}
			p V = N k_B T .
		\end{equation}
		Assuming a more or less constant density we infer
		\begin{equation}
			\bmath{\nabla} p \approx n k_B \bmath{\nabla} T .
		\end{equation}
		Knowing that the source of a heat flux corresponds to the time evolution of pressure and using Eq. (\ref{EQbasicheatflux}) we obtain
		\begin{equation}
			\frac{\partial p}{\partial t} \sim \bmath{\nabla} \cdot Q ~~ \rightarrow ~~ n k_B \frac{\partial T}{\partial t} \sim \bmath{\nabla} \cdot \lp \kappa \bmath{\nabla} T \rp .
		\end{equation}
		Discretising the derivatives by typical length and time scales we get for the conduction coefficient
		\begin{equation}
			\kappa \sim \frac{l^2}{\tau} \cdot n k_B \sim D \cdot n k_B ,
			\label{EQconductiondiffusionrelation}
		\end{equation}
		where we can identify the diffusion coefficient $D$. According to this relation, the two coefficients behave similarly and we can therefore apply the following scaling relations on an implementation of anisotropic thermal conduction.
		
		\subsubsection{Coefficient proportional to $B^{-2}$}			
			
			At first, we connect the mean free path and collision time via the particle's velocity $\lambda \approx v \tau$.
			A typical diffusion coefficient of units $cm^2 ~ s^{-1}$ can be defined as
			\begin{equation}
				D \approx \frac{\lambda^2}{\tau} \approx \lambda v \approx v^2 \tau .
			\end{equation}
			Since particle movement parallel to the magnetic field is not restricted, the diffusion along the field lines should not be affected, which corresponds to $D_\parallel = D$.
			
			We assume that motion of particles perpendicular to magnetic field lines is only possible by jumps between cyclotron trajectories, which results in a diffusion coefficient like
			\begin{equation}
				D_\perp \approx \frac{v^2}{\omega_g^2 \tau} \approx \frac{\lambda v}{w_g^2 \tau^2} .
			\end{equation}
			Therefore, the relation between the two coefficients is
			\begin{equation}
				\frac{D_\perp}{D_\parallel} \approx \frac{1}{\omega_g^2 \tau^2} \propto B^{-2} .
			\end{equation}
			This is however only valid if $\omega_g \tau >> 1$, ergo if the gyroradius is much smaller than the mean free path. In other words, we need the magnetic field to impose a notable restriction onto the electrons movement. In the regime of $\omega_g \tau \sim 1$ we have to change the relation in order to ensure $D_\perp \le D_\parallel$:
			\begin{equation}
				\frac{D_\perp}{D_\parallel} \approx \frac{1}{1 + \omega_g^2 \tau^2} .
				\label{EQdiffusionbsquared}
			\end{equation}
			To evaluate this relation for a given system we require the collision time or the corresponding frequency
			\begin{equation}
				\frac{1}{\tau} = \nu = \frac{\omega_{pl}}{n \lambda_D^3} ,
				\label{EQcollisionfrequency}
			\end{equation}
			with the plasma frequency
			\begin{equation}
				\omega_{pl} = \sqrt{\frac{4 \pi n e^2}{m}}
			\end{equation}
			and the Debye length
			\begin{equation}
				\lambda_D = \sqrt{\frac{k_B T}{4 \pi n e^2}} .
			\end{equation}
			Putting these relations together we finally obtain
			\begin{equation}
				\frac{1}{\omega_g \tau} = \frac{16 \pi^2 c \, e^3 \sqrt{m} \, n}{\lp k_B T \rp^{3/2} B} \approx 10^{-5} ~ \frac{n}{T^{3/2} B} ~ \mathrm{cm}^3 \, \mathrm{K}^{3/2} \, \mathrm{G} .
			\end{equation}
			
			To check the order of magnitude of the fraction of perpendicular and parallel diffusion coefficient, we use typical values for the magnetic field strength, temperature and density in galaxy clusters (see section \ref{introduction}). Eq. (\ref{EQdiffusionbsquared}) results in a factor of $D_\perp / D_\parallel \sim 10^{-28}$ which means that conduction perpendicular to the magnetic field is be typically extremely suppressed in the ICM.
		
		\subsubsection{Coefficient proportional to $B^{-1}$}
		
			However, these relations are only phenomenological estimates.Additionally, perpendicular diffusion is overlayed with turbulence transport processes, which are extremely difficult to describe. However, laboratory experiments show, that the scaling with the magnetic field effectively changes from $B^{-2}$ to $B^{-1}$, which characterises so called Bohm diffusion. \citep{Guthrie1949} According to the calculations above we construct a scaling relation for this kind of behaviour of
			\begin{equation}
				D_\perp \approx \frac{v^2}{\omega_g} \approx \frac{k_B T c}{e B} .
			\end{equation}
			We assume an electrons movement with their thermal speed $v \approx \sqrt{k_B T / m}$ and neglect further influences for example by plasma instabilities. We get in total
			\begin{equation}
				\frac{D_\perp}{D_\parallel} \approx \frac{1}{\omega_g \tau} ,\label{EQdiffusionblinear}
			\end{equation}
			which allows a much stronger diffusion orthogonal to the magnetic field lines for typical values.
			A more detailed analysis with similar results is given for example in \cite{Golant1980}.
			
			For highly tangled magnetic fields \cite{Pistinner1996} discuss if the coherence length should replace the gyroradius. However, they find that this assumption is wrong. This matches the considerations of \cite{Rosner1989} who present that tangled magnetic fields do not suppress thermal conduction very strongly, despite general believe. Their results state a reduction factor of $\left< \cos \delta \theta \right>^2$ which is an average over the local angles between magnetic field lines and the temperature gradient. We briefly analyse this behaviour for totally random magnetic field configurations in section \ref{temperaturestep}.
			
			Summing up, thermal conduction perpendicular to magnetic fields lines with reasonable field strengths is in general almost totally suppressed. When we come into a regime where we need to apply scaling relations regarding the magnetic fields, the ratio $\kappa_\perp / \kappa_\parallel$ scales either like $B^{-2}$ or $B^{-1}$.
			
		\subsubsection{Coefficient for the cross product term}
			\label{physicalhall}
			
			\nr{Finally we come to the last term of eq. \ref{EQconductionbraginskii} which is a bit different then the other two contributions to thermal conduction. So far we have only talked about collisional thermal conduction. As the name states, heat is exchanged microscopically due to collisions of particles. This last term arises however from a different origin, namely the Hall effect. This term handles heat transport in the third spatial direction, perpendicular to both the temperature gradient and the magnetic field. According to \cite{Braginskii1965} it scales also linear with the magnetic field strength, amounting in a similar description of the coefficient as presented in the previous section. We will not go into further detail here, since we will later show that this term plays no role in our discretisation scheme (see section \ref{vanishhall}).}
			
		\subsection{The anisotropic conduction equation}
		
		We finalize our considerations and derive the anisotropic conduction equation. In principle, there are different possible approaches. We briefly repeat the immediate requirements for the resulting scheme:
		\begin{itemize}
			\item Unchanged isotropic conduction if the magnetic field is parallel to the temperature gradient,
			\item Strong suppression of energy transfer via conduction if the magnetic field is perpendicular to the temperature gradient,
			\item Scaling of the suppression factor inverse with the magnetic field strength to some power.
		\end{itemize}
		A simplistic approach to fulfil the first two requirements, however not the third, can be taken by multiplying the projection of the magnetic field onto the temperature gradient to the conduction coefficient
		\begin{equation}
		\bmath{Q} = - \kappa \frac{\bmath{\nabla} T \cdot \bmath{B}}{\abs{\bmath{\nabla} T \cdot \bmath{B}}} \bmath{\nabla} T =: - \lp \kappa \cos \theta \rp \bmath{\nabla} T .
		\end{equation}
		This approach has the advantage that it requires barely any change in the existing numerical scheme presented in \cite{Jubelgas2004} and does not cost much additional computation time. It can be regarded as motivation for the 1/3 suppression factor of Spitzer conduction in earlier simulations including only isotropic conduction. However, we have no possibility to introduce a scaling of the suppression dependent on the actual strength of the magnetic field. We therefore would have to assume a sufficiently strong magnetic field, which can not be guaranteed in the whole computational domain at all times. Problems arise especially in combination with a magnetic seeding mechanism when no initial magnetic field is present.
		
		Instead we follow a different derivation by splitting up the conduction equation into a part parallel and a part perpendicular to the magnetic field and assign different conduction coefficients to both parts
		\begin{equation}
		\frac{du}{dt} = \frac{1}{\rho} \bmath{\nabla} \cdot \lb \kappa_\parallel \lp \bmath{\hat{B}} \cdot \bmath{\nabla} T \rp \bmath{\hat{B}} + \kappa_\perp \lp \bmath{\nabla} T - \lp \bmath{\hat{B}} \cdot \bmath{\nabla} T \rp \bmath{\hat{B}} \rp \rb .
		\end{equation}
		With $\bmath{\hat{B}}$ being the normalised magnetic field vector. \nr{For the moment we focus only on the collisional terms and come back to the hall term, later.}
		
		Please note, that we can easily regain the isotropic equation from this by setting $\kappa_\parallel = \kappa_\perp$ or $\bmath{\hat{B}} = \bmath{0}$. Plugging in the relation between parallel and perpendicular diffusion we derived earlier, it can be seen that all of our requirements are fulfilled.
		
		We reshuffle the terms for better handling.
		\begin{equation}
		\frac{du}{dt} = \frac{1}{\rho} \bmath{\nabla} \cdot \lb \lp \kappa_\parallel - \kappa_\perp \rp \lp \bmath{\hat{B}} \cdot \bmath{\nabla} T \rp \bmath{\hat{B}} + \kappa_\perp \bmath{\nabla} T \rb
		\label{EQconductionfinalsecondapproach}
		\end{equation}
		
		From section \ref{anisotropicthermalconduction} we know that within galaxy clusters mainly $\kappa_\perp \ll \kappa_\parallel$. However, we can not simply neglect the second term along the temperature gradient. Comparing the absolute values of the two terms we see, that except of $\kappa_\perp \ll \kappa_\parallel$ the first term contains a $\cos \theta$ which can be arbitrarily small and make both terms comparable in magnitude. If the magnetic field and the energy gradient are almost totally perpendicular, the second term dominates and can not be neglected. For further reference on how exemplary previous work (mostly on galaxy clusters) of the past 15 years has handled the different terms of thermal conduction please see \cite{Ruderman2000, Dolag2004, Schekochihin2008, Rasera2008, Parrish2009, Sharma2010, ZuHone2013, Suzuki2013, Komarov2014, ZuHone2015, Dubois2015, Kannan2015, Yang2015}. 
\section{Numerical implementation}
	\label{numericalimplementation}

	In this section we derive our numerical representation of anisotropic thermal conduction.
	
	Before transforming Eq. (\ref{EQconductionfinalsecondapproach}) into SPH formalism we note that the second term can be handled similar to the isotropic implementation just with a different coefficient. For details we refer to \cite{Jubelgas2004}. Here we discuss only on the first term.

	\subsection{A fully consistent derivation}
		\label{finalnumerics}
		
		Initially, the term suggests a split-up of the calculation of temperature gradient and divergence. However, this requires a large amount of additional computation time, since it needs an additional SPH loop, but also leads to further numerical errors due to the effective second kernel derivative introduced (see \citealt{Brookshaw1985}). Furthermore, chaining SPH discretisations can introduce strongly growing numerical errors and should be avoided.
		
		Instead, we derive a consistent formulation with only one SPH loop following the example of \cite{Petkova2009}, who developed an SPH scheme for a similar diffusion equation in radiative transfer. In the following calculations latin indices like $i$, $j$ and $k$ always denote particles while greek indices like $\alpha$, $\beta$ indicate components of tensors.
		
		Before we start discretising the modified conduction equation, we have to find a better estimate for mixed second derivatives. The derivation is at first similar to the one presented by \cite{Jubelgas2004}, but gets more complicated since we also need mixed derivatives.
		
		Consider an arbitrary quantity $Q$ at $\xj$ which we expand around $\xi$
		\begin{equation}\begin{array}{ll}
		Q \lp \xj \rp \approx & Q \lp \xi \rp + \at{\bmath{\nabla} Q}{\xi} \xij + \\
		& \frac{1}{2} \sum\limits_{\alpha \beta} \at{\frac{\partial^2 Q}{\partial x_\alpha \partial x_\beta}}{\xi} \lp \xij \rp_\alpha \lp \xij \rp_\beta + \mathcal{O} \lp \lp \xij \rp^3 \rp .
		\label{EQqnumericstaylor}
		\end{array}\end{equation}
		With the distance vector \mbox{$\xij = \xi - \xj$}.
		
		We multiply both sides by $\frac{\lp \xij \rp_\gamma}{\abs{\xij}^2}  \frac{\partial \Wij}{\partial \lp \xi \rp_\delta}$ and integrate over $\int d^3 \xj$.
		
		The first order term vanishes due to antisymmetry of the integrand and we solve for the second order term. More detailed calculations are presented in appendix \ref{appendixsecondorder}. Assuming Q is a second order tensor quantity, we can rewrite the equation to
		\begin{equation}
		\sum\limits_{\alpha, \beta} \frac{\partial^2 \lp Q_i \rp_{\alpha \beta}}{\partial x_\alpha \partial x_\beta} = 2 \int d^3 \bmath{x}_j ~ \frac{\xij^{\; \intercal} \lb \bmath{\tilde{Q}}_j - \bmath{\tilde{Q}}_i \rb \bmath{\nabla}_i \Wij}{\abs{\xij}^2} \label{EQmixedsecondderivativeestimatefirst}
		\end{equation}
		and
		\begin{equation}
		\sum\limits_{\alpha, \beta} \lp Q_i \rp_{\alpha \beta} \frac{\partial^2 T_i}{\partial x_\alpha \partial x_\beta} = 2 \int d^3 \bmath{x}_j ~ \frac{\xij^{\; \intercal} \bmath{\tilde{Q}}_i \lb T_j - T_i \rb \bmath{\nabla}_i \Wij}{\abs{\xij}^2} \label{EQmixedsecondderivativeestimatesecond}
		\end{equation}
		with the substituted tensor
		\begin{equation}
			\tilde{\mathbfss{Q}} = \frac{5}{2} \mathbfss{Q} - \frac{1}{2} \tr{\mathbfss{Q}} \unity .
			\label{EQtildevariabledef}
		\end{equation}
		This is a very compact and neat formulation and we use \cite{Jubelgas2004} to check this formula for consistency.
		
		Consider $\mathbfss{Q} = Q \cdot \mathbfss{1}$. Then we get
		\begin{equation}
			\bmath{\tilde{Q}} = \frac{5}{2} \cdot Q \cdot \bmath{1} - \frac{1}{2} \cdot Q \cdot 3 \cdot \bmath{1} = Q \cdot \bmath{1}
		\end{equation}
		Putting this into Eq. (\ref{EQmixedsecondderivativeestimatefirst}) or (\ref{EQmixedsecondderivativeestimatesecond}) we recover the result which can be obtained for the isotropic implementation, where only non-mixed second derivatives are needed:
		\begin{equation}
			\sum_{\alpha} \frac{\partial^2 Q_i}{\partial x_\alpha^2} = 2 \int d^3\xj \lp Q_j - Q_i \rp \frac{\xij^{\; \intercal} \cdot \bmath{\nabla}_i \Wij}{\abs{\xij}^2} .
		\end{equation}
		
		Before we further analyse the properties of these approximation formulas let us at first review our basic equation.
		
		As previously mentioned we consider only the part of Eq. (\ref{EQconductionfinalsecondapproach}) parallel to the magnetic field (the first term). The term conducting along the temperature gradient can be handled isotropically, which is described in \cite{Jubelgas2004}.
		
		We start by writing the equation in component form:
		\begin{equation}
		\at{\frac{du}{dt}}{1st} = \frac{1}{\rho} \sum\limits_{\alpha, \beta} \frac{\partial}{\partial x_\alpha} \lb \lp \kappa_\parallel - \kappa_\perp \rp \hat{B}_\alpha \hat{B}_\beta \frac{\partial}{\partial x_\beta} T \rb . \label{EQanisoconductioncomponents}
		\end{equation}
		Furthermore, we define the components of a tensor $\bmath{A}$ as
		\begin{equation}
			\Aab := \lp \kappa_\parallel - \kappa_\perp \rp \hat{B}_\alpha \hat{B}_\beta .
		\end{equation}
				
		Next, we write the equation only in terms of mixed second derivatives:
		\begin{equation}
			\at{\frac{du}{dt}}{1st} = \frac{1}{2 \rho} \sum\limits_{\alpha, \beta} \lp \frac{\partial^2 \Aab T}{\partial x_\alpha \partial x_\beta} - T \frac{\partial^2 \Aab}{\partial x_\alpha \partial x_\beta} + \Aab \frac{\partial^2 T}{\partial x_\alpha \partial x_\beta} \rp .
			\label{EQconductionanisosecondderivatives}
		\end{equation}
		
		Now we use Eq. (\ref{EQmixedsecondderivativeestimatefirst}) and (\ref{EQmixedsecondderivativeestimatesecond}) to estimate the second derivatives in Eq. (\ref{EQconductionanisosecondderivatives}). Re-factoring the terms leads to a compact expression for particle \textit{i}:
		\begin{equation}
			\at{\frac{du_i}{dt}}{1st} = \frac{1}{\rho_i} \int d^3 \xj  ~ \xij^{\; \intercal} \lb \frac{\lp \tilde{\mathbfss{A}}_j + \tilde{\mathbfss{A}}_i \rp \lp T_j - T_i \rp}{\abs{\xij}^2} \rb \bmath{\nabla}_i \Wij .
		\end{equation}
		Finally, we discretise the integral and rewrite the temperature to specific internal energy:
		\begin{equation}\begin{array}{ll}
			\at{\frac{du_i}{dt}}{1st} = & \frac{\mu \lp \gamma - 1 \rp}{k_B \cdot \rho_i} \cdot\\
			& \ngbsum \frac{m_j}{\rho_j} \cdot \xij^{\; \intercal} \lb \frac{\lp \tilde{\mathbfss{A}}_j + \tilde{\mathbfss{A}}_i \rp \lp u_j - u_i \rp}{\abs{\xij}^2} \rb \bmath{\nabla}_i \Wij
			\label{EQansiofinalfinal}
		\end{array}\end{equation}
		with the mean molecular mass $\mu$ and the adiabatic index $\gamma$. This equation allows us to calculate the effects of anisotropic conduction without an additional SPH loop.
		
		One convenient property is that we managed to generate the term $\lp T_j - T_i \rp$ like in the isotropic conduction case. This ensures only conduction if the temperatures of two particles differ and the sign takes care of the heat flux' direction. 
		
	\subsection{Ensuring the 2nd law of thermodynamics}
		
		There might still be a problem with this approximative formula. To ensure the correct flow of internal energy from hot to cold (according to the second law of thermodynamics) the tensor $\lp \tilde{\mathbfss{A}}_j + \tilde{\mathbfss{A}}_i \rp$ must be positive definite. However, from the definition of a variable with tilde (Eq. (\ref{EQtildevariabledef})) we see, that this tensor does not necessarily fulfil this condition. For a very anisotropic setup heat flows in the wrong direction. In addition to a violation of classical thermodynamics this \nr{can lead to numerical instabilities depending on the solved used (please see section \ref{conjugategradient} for further details).} To overcome this problem we have basically three options which are both artificial and therefore might negatively influence onto our discretisation formula in general:
		\begin{enumerate}
			\item Implement a limiter in the code, which checks for non physical heat flows.
			\item \nr{Check if a configuration leads to wrong flux and prevent anisotropic conduction}
			\item Change the tensor to a more isotropic version, which is always positive definite.
		\end{enumerate}
		\nr{Flux limiters in a sense that each heat flux is decreased by a certain amount scaling with it's initial amount are hard to judge. There exist several approaches as shown by \cite{Petkova2009,Koerner2014} but usually one has to settle down with an empirically found formulation. It is hard to determine whether a given limiter is the best one for a certain problem. However, we actually already employ a physically motivated limiter in our code: the consideration of a saturated heat flux for low density plasmas given by Eq. \ref{EQsaturatedflux}. While this does not guarantee us to prevent all non physical heat flux it turns out to do a very good job in our galaxy cluster simulations keeping the convergence of our solver fast and well defined.
		
		The second option seems to be a rather artificial one. We can determine the angle between the temperature gradient and the magnetic field for each SPH particle and clearly detect, whenever heat would flow in the wrong direction. In this case we could either forbid all conduction or simply fall back to the isotropic formulation. Being a mixture of anisotropic and isotropic or even totally suppressed conduction, this kind of algorithm would be extremely hard to control and render any results of cosmological simulations incomprehensible.
		
		\cite{Petkova2009} also propose the third option, namely to add an isotropic component to the anisotropic tensor in order to prevent temperature flowing from cold to warm regions. We already have a pure isotropic component which is however proportional to $\kappa_\perp$ and it is not clear if this is already sufficient.
	
		While the first option is hardly applicable in our tests, we will go on by comparing the \textit{fully anisotropic} formulation with an \textit{isotropised} version in our tests and cluster simulations in the next sections. In order to ensure that option 3 works, we add an artificial isotropic component and replace the tensor $\tilde{\mathbfss{A}}$ by}
		\begin{equation}
			\tilde{\mathbfss{A}} \rightarrow \alpha \tilde{\mathbfss{A}} + \frac{1}{3} \lp 1 - \alpha \rp \tr{\tilde{\mathbfss{A}}} \bmath{1} .
		\end{equation}
		Calculations carried out by \cite{Petkova2009} show that we need to set $\alpha \ge \frac{2}{5}$. We use the minimum value to prevent a large error in the estimate. This leads to $\tilde{\mathbfss{A}} \rightarrow \mathbfss{A}$, which is computationally very cheap since we have to compute $\mathbfss{A}$ for each particle, anyway. We call this formulation the \textit{isotropised} discretisation.
		
		We can check, that $\mathbfss{A}$ itself is positive definite by diagonalising it:
		\begin{equation}
			\mbox{diag} \lp \mathbfss{A} \rp \propto \mbox{diag} \lp  \hat{B} \otimes \hat{B} \rp = \lp \begin{array}{ccc} 0&0&0\\0&0&0\\0&0&1 \end{array} \rp .
		\end{equation}
		
		This isotropic term added to the conduction matrix helps to remove non-physical heat flux; however, it does not come straight forward from our derivation of the conduction equation.
		It is artificially added and presents itself as an offset to the direct derivation.
		Therefore, we can not expect that this isotropised version of the heat flux equation will still behave exactly as the undiscretised equation dictates.
		To which degree this adjusted formulation is better or worse than the fully anisotropic description remains to be investigated by the test cases.
		
		Similar problems of non physical heat fluxes arise also in grid code solutions and are not an intrinsic problem of SPH formulations \citep[see e.g.][]{Sharma2007}.
		
	\subsection{Discretising the hall term}
		\label{vanishhall}
		
		\nr{So far we have left out the cross product term of eq. \ref{EQconductionbraginskii} in our discretisation. However, this can be easily done in the same fashion as before. By writing the equation component wise we can again define a 3x3 tensor $\mathbfss{A}_\Lambda$:
		\begin{equation}
			A_{\Lambda ~ \alpha \beta} = \kappa_\Lambda \hat{B}_\gamma \epsilon_{\alpha \gamma \beta},
		\end{equation}
		which is then differentiated as eq. \ref{EQanisoconductioncomponents}, calculating
		\begin{equation}
			\frac{\partial}{\partial x_\alpha} \lp A_{\Lambda ~ \alpha \beta} \frac{\partial}{\partial x_\beta} T \rp
		\end{equation}
		This allows us to follow the same path of the derivation as before.
		
		However, one crucial difference is not to be overlooked: While the matrix $\mathbfss{A}$ is symmetric, $\mathbfss{A}_\Lambda$ is antisymmetric due to the epsilon tensor in it's definition. As we see in eq. \ref{EQansiofinalfinal}, the matrix is basically multiplied with the position difference vector of two particles from both sides, since the kernel gradient also points in that direction. But, if an antisymmetric matrix is multiplied by the same vector from both sides, the result is zero. This can be easily shown by changing the order of the terms:
		\begin{equation}\begin{array}{ll}
		\xij \mathbfss{A}_\Lambda \xij &= x_{ij \alpha} \cdot \kappa_\Lambda \hat{B}_\gamma \epsilon_{\alpha \gamma \beta} \cdot x_{ij \beta}\\
		&= - \kappa_\Lambda \cdot x_{ij \alpha} x_{ij \beta} \epsilon_{\alpha \beta \gamma} \cdot \hat{B}_\gamma\\
		&= - \kappa_\Lambda \lp \xij \times \xij \rp \hat{B} = 0
		\end{array}\end{equation}
		Therefore, the hall term vanishes in our discretisation and in all other discretisations with the same property.
		}
				
	\subsection{Solving the differential equation}
		\label{conjugategradient}
		
		Finally, we address the time integration of the resulting equation.
		
		\cite{Jubelgas2004} show how to apply a symmetry enforcing finite difference scheme. This is a fairly simple and computationally cheap approach, however, they conclude that a kernel averaging of the temperature is required to suppress the effects of small-scale noise in the temperature distribution. Therefore, an additional SPH loop is required, which greatly increases the computational cost.
		
		In contrast, \cite{Petkova2009} considered an implicit integration scheme. That requires again an additional SPH loop but has the advantage of much more accurate results for larger conduction time steps, therefore reducing the computational cost. They chose the so called conjugate gradient (CG) method which \nr{we discuss in the following subsection followed by an analysis of a possibly more stable alternative.}.
		
		\subsubsection{The conjugate gradient}
			\nr{The CG solver} is basically an algorithm to solve a matrix inversion problem. Instead of fully inverting the matrix, it is also \nr{often} used as an iterative approximation method with very good convergence properties. \nr{In our case such an iterative approach is mandatory, since the matrix we are dealing with is of dimension particle number squared. Inverting this matrix explicitly would consume way to much time to be a viable. Basically the algorithm calculates the direction in which it has to iterate in order to monotonically approach the correct solution with each iteration step being weighted by the residual of the previous one. Such an iterative algorithm requires an initial guess for the solution, for which we plug in the current state of the system. Assuming that thermal conduction only changes energies on small scales within reasonable time steps, this approach results in a fast convergence rate for the algorithm.} A detailed discussion of the algorithmic properties is for example given by \cite{Saad2000}.
			
			Due to the advantages in overall computational cost and precision we want to use the conjugate gradient method to solve the anisotropic conduction equation. At first, we need to ensure, that our equation suffices the requirements of the CG solver. To discuss the properties consider the following equation which we want to solve for the vector $\bmath{x}$:
			\begin{equation}
				\mathbfss{C} \cdot \bmath{x} = \bmath{b} .
			\end{equation}
			For the algorithm to succeed, we place a few constraints on the matrix $\mathbfss{C}$: It needs to be real, symmetric and positive definite.
			Because the equations do not contain imaginary parts, the first condition is always fulfilled.
			
			Symmetry corresponds to conservation of energy \nr{since each matrix entry poses the heat flow between two particles}, which should always be fulfilled for an energy transport scheme. If this was not the case self-consistently from the derivation, we would have to symmetrise the result afterwards.\footnote{Please note that we would have to consider the real internal energy, hence the equations for $u_i$ derived above have to be multiplied by $m_i$.} We see if this property is fulfilled after writing down the equation explicitly as a matrix inversion problem.
			
			The positive definiteness can be argued as follows: In the continuous limit the matrix becomes diagonal. Positive definite for a symmetric and real matrix means that the eigenvalues are positive. In our case this corresponds to heat being transported only anti-parallel to the temperature gradient following the 2nd law of thermodynamics. We argued about positive definiteness already in section \ref{finalnumerics}: The fully anisotropic formulation can violate this condition, which can therefore lead to non-physical heat flows as well as numerical instabilities, since the CG method in principle requires it to be given. The isotropised version is constructed such that it definitely fulfils positive definiteness.
			
			Now we show how to write Eq. (\ref{EQansiofinalfinal}) in CG formalism. Discretising the timestep using
			\begin{equation}
				\frac{du_i}{dt} \rightarrow \frac{\Delta u_i}{\Delta t} = \frac{u_i^{n+1} - u_i^n}{\Delta t}
			\end{equation}
			we get for the part along the magnetic field lines
			\begin{equation}
				u_i^{n+1} = u_i^n + \ngbsum c_{ij} \lp u_i^{n+1} - u_j^{n+1} \rp
			\end{equation}
			with 
			\begin{equation}
				c_{ij} = - \frac{\lp \gamma - 1 \rp \mu}{k_B} \cdot \frac{m_j \Delta t}{\rho_i \rho_j} \cdot \frac{\xij^{\; \intercal}}{\abs{\xij}^2} \lp \tilde{\mathbfss{A}}_i + \tilde{\mathbfss{A}}_j \rp \bmath{\nabla}_i \Wij .
			\end{equation}
			
			We can then write this as the matrix equation $\mathbfss{C} \cdot \bmath{x} = \bmath{b}$ with:
			\begin{itemize}
				\item $C_{ij} := \delta_{ij} \lp 1 - \sum_k c_{ik} \rp + c_{ij}$
				\item $x_j := u_j^{n+1}$
				\item $b_i := u_i^n$
			\end{itemize}
			Now, we check again if the energy is conserved properly (except for numerical errors). $\tilde{\mathbfss{A}}$ and therefore $\bmath{c}$ and $\mathbfss{C}$ are symmetric, which is exactly the property we identified with energy conservation. For the isotropised version we get the same equations just without the tilde above each $\mathbfss{A}$ and therefore same argumentation holds.
			
		\subsubsection{An improved approach: The bi-CGStab}
			\nr{Since it is worrisome that the conjugate gradient solver may be numerically unstable for a non positive definite matrix we also propose a slightly different algorithm: The bi-Conjugate Gradient. Geometrically speaking this solver does not only converge monotonically along one direction given by the gradient as the CG does, but also iterates along a second vector. This ensures that at a saddle point where the gradient can not be determined properly, the algorithm does not simply get stuck before convergence. There exist several flavours of the bi-CG like for example the Conjugate Gradient-Squared \citep{Sonneveld1989} or the bi-Conjugate Gradient Stabilised \citep{Vorst1992}.
				
			We propose using the latter algorithm since it poses a very good way around the difficulties with just a few changes required. Since in each iteration two vectors are followed, two matrix-vector multiplications are needed which a priori doubles the computational cost in comparison to the CG. However, the bi-CGStab is supposed to have rather smooth convergence properties (in contrast to other flavours of the biCG) and even uses both operations to increase the convergence rate. In the case of a positive definite matrix it even falls back to the CG, only with two steps in one iteration meaning that bottom line no additional computational cost has been added.
			
			Of course there exist also totally different algorithms to handle the matrix inversion problem, the most popular being probably GMRES \citep{Saad1986}. However, these are also not guaranteed to perform better with faster convergence and therefore we prefer to use a conjugate gradient style of method in order to maintain the maximum possible backwards compatibility in our code.}
			
		\subsubsection{Further improvements: preconditioning}
			\nr{Independent of the solver, as long as we treat the problem as a matrix inversion, there exists the possibility to speed up convergence of the iteration by applying a preconditioner to the matrix. Since our matrix contains the particle-particle interaction terms it has a rather sparse pattern, limited by the amount of neighbours. Therefore, we do not save the whole matrix in memory but calculate the elements on demand. Moreover, the sparsity pattern is highly dependent on the particle ordering by the SPH tree. This makes it highly non trivial to find a good preconditioner which always helps to speed up convergence. Since our tests reveal, that convergence is not an issue for us we refrain from implementing a preconditioning matrix here and keep the possibility in mind should the need arise. For more literature on this topic we like to refer to the references given by \cite{Vorst1992}.}

\section{Tests for the new code}
	\label{tests}

	Next, we carry out several tests for the different implementations. We use rather simple test cases for which we can verify the behaviour analytically, before we apply it to a physically challenging problem like galaxy cluster evolution.
	
	For all of the following tests we use initial conditions with "glass-like" particle distributions. Therefore, we rule out any alignment effects which arise by the definition of a grid. Even if the test setups could be done in one or two dimensions, we perform all tests in a fully three dimensional set-up.
	
	Furthermore, we run the simulations with gas only and disable any accelerations on the SPH particles which would come from self-gravity or the MHD equations. With this approach, we ensure that hydrodynamical properties like the density and the internal energy are computed correctly in their respective SPH loops, but we evolve only the conduction equation to thoroughly test the behaviour of our implementations. \nr{Additionally, we keep the heat flux limiter due off to gain a better understanding of how the code behaves.}
	
	We always start by describing a test case and the derivation of an analytic solution. We are only be able to derive an analytic solution for a constant conduction coefficient, which we enforce in our code for the test problems instead of using Spitzer conduction. Afterwards we show the behaviour of the existing code (i.e. isotropic conduction) with a reference run and further present our results with the new anisotropic approaches (fully anisotropic and isotropised).
	
	Finally, we present a more complicated test, where we allow a temperature dependent Spitzer conduction and check the influence of different prescriptions for perpendicular suppression.

	\subsection{Temperature step problem}
		\label{temperaturestep}
		
		At first, we reproduce the first test of \cite{Jubelgas2004} and slightly modify it, so that we can apply it to the new anisotropic conduction implementation.
		The basic idea is to set-up a temperature step and let the particles exchange heat. We fix the particle positions (and also the magnetic field, which we add later) and therefore only evolve the conduction equation. Also considering a fixed conduction coefficient instead of Spitzer conduction we can pull the $\kappa$ out of the divergence and get
		\begin{equation}
			\frac{du}{dt} = \frac{\kappa}{\rho} \Delta T . \label{EQsimplifiedconductioneq}
		\end{equation}
		This simplified conduction equation can be solved analytically (depending on the initial conditions) and we compare to the simulation results. We assume a gas with constant density and use
		\begin{equation}
			u = c_v \cdot T .
		\end{equation}
		with the specific heat capacity $c_v$. We rewrite Eq. (\ref{EQsimplifiedconductioneq}) to
		\begin{equation}
			\frac{du}{dt} = \alpha \cdot \Delta u \label{EQsimplifiedconductioneqshorter}
		\end{equation}
		with the so called thermal diffusivity $\alpha = \kappa/c_v\rho = \mathrm{const}$, which is simply a diffusion coefficient, as discussed in section \ref{anisotropicthermalconduction}.
		For this temperature step problem it is sufficient to solve the equation in one dimension. The more general solution can be inferred later and basically differs only in some pre-factors. Following \cite{Jubelgas2004} and \cite{Landau2007} this equation can be solved through Fourier transformation. For details please see appendix \ref{appendixtemperaturestep}.
		
		We describe the initial internal energy distribution with the following step function:
		\renewcommand{\arraystretch}{1.3}
		\begin{equation}
			u_0(x') =  
			\left\{\begin{array}{ll}
			u_0 - \frac{\Delta u}{2} & \mbox{for } x' < x_m\\
			u_0 + \frac{\Delta u}{2} & \mbox{for } x' > x_m .
			\end{array}\right.
		\end{equation}
		\renewcommand{\arraystretch}{1.0}
		with $x_m$ being the position of the temperature step, $\Delta u$ the height and $u_0$ the mean value. We get in total
		\begin{equation}
			u(t,x) = u_0 + \frac{\Delta u}{2} \cdot \erf \lp \frac{x-x_m}{2\sqrt{\alpha t}} \rp .
			\label{EQtemperaturestepfinal}
		\end{equation}
		
\begin{figure}
			\centering
			\includegraphics[angle=270,width=0.85\linewidth]{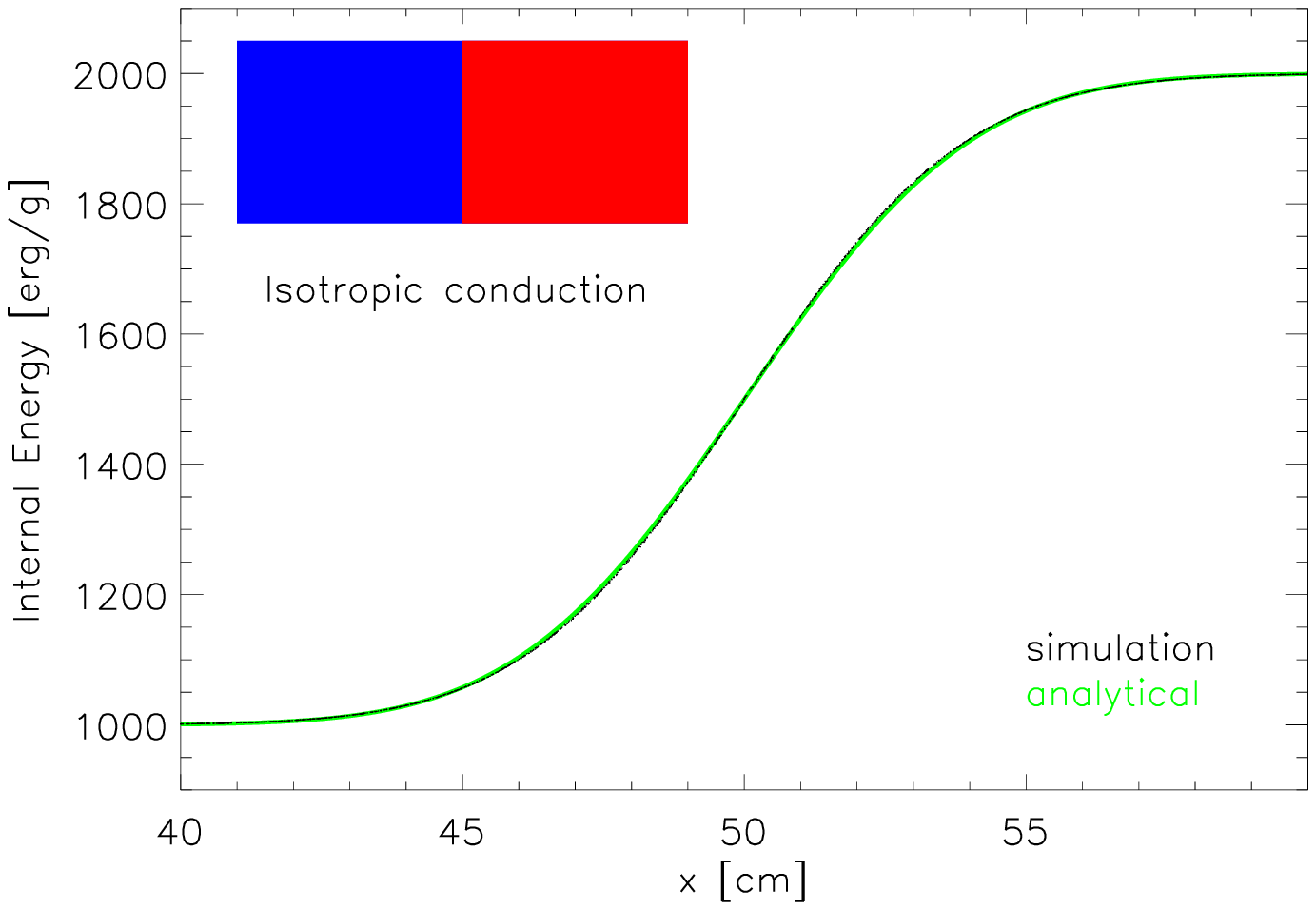}
			\caption{First conduction test: One dimensional temperature step without magnetic field. The green line is the analytical solution (Eq. (\ref{EQtemperaturestepfinal})), the black dots are SPH particles. Both solutions match very well.}
			\label{FIGconduction1a}
		\end{figure}

		At first, we cross check our calculations with the existing implementation of isotropic conduction. The result is shown in Fig. \ref{FIGconduction1a}. The SPH particles are directly plotted as black points without any binning or additional smoothing. The result matches well with the analytic solution. Therefore, the existing implementation works even for sudden temperature jumps.
		
		\subsubsection{Diagonal magnetic field}
		
			The next step is to include a magnetic field into the test problem to check the new anisotropic implementation. For simplicity we keep the magnetic field fixed. We introduce a homogeneous field in direction $\lp x, y, z \rp = \lp 1/\sqrt{2}, 1/\sqrt{2}, 0 \rp$. Hence, there is an angle of $45^\circ$ between the $\bmath{B}$-field and the energy gradient, which is in our set-up parallel to the $x$-axis.
			
			\begin{figure}
				\centering
				\includegraphics[width=\linewidth]{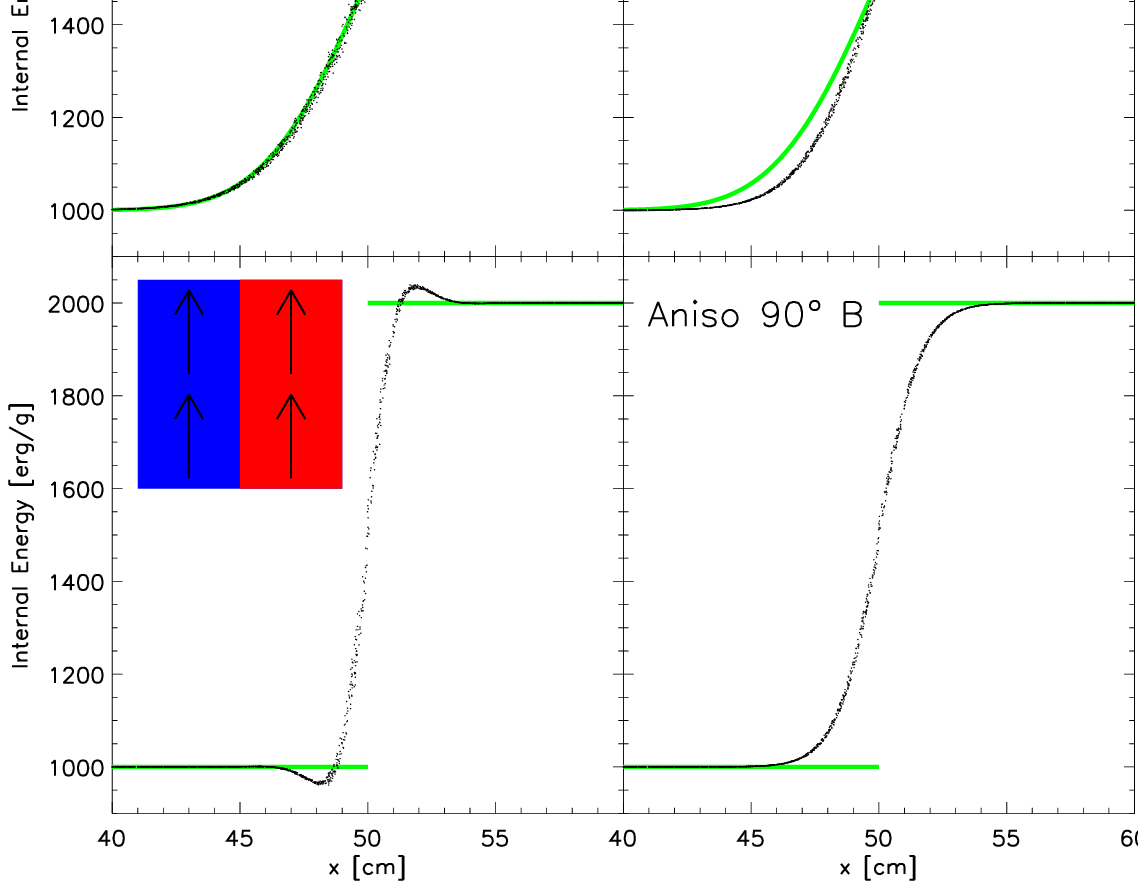}
				\caption{The temperature step problem, simulated with the fully anisotropic implementation left and the isotropised version right for a diagonal (upper row), parallel (middle row) and vertical magnetic field (lower row). All plots are made at the same simulation time.}
				\label{FIGconduction1all}
			\end{figure}
			
			The results of the fully anisotropic conduction version at the same simulation time as before are plotted in Fig. \ref{FIGconduction1all} in the top left box. This implementation reproduces the analytic solution quite well; however, we get more scatter than in the run with isotropic conduction. It is still not yet clear what the origin for this noise exactly is, but for a cosmological simulation this is of less importance, since thermal conduction is not the dominating effect modifying and the scatter is smoothed out. Probably, this formulation does not ensure a positive definite transport matrix, which induces errors into the conjugate gradient solver. However, we see that for a $45^\circ$ magnetic field we do not need an artificial isotropisation to obtain a stable solution.
			
			The top right panel shows the results using the isotropised formulation. Clearly, we never get the exact analytic result, since the isotropisation is artificially added into the numerics, but our result is close to the real solution. In contrast to the fully anisotropic formulation we get less scatter since the anisotropic part of the equation is mixed with an isotropic component and therefore has a weaker effect. Furthermore, we ensure positive definiteness of the transport matrix which guarantees stability of the algorithm.
			
		\subsubsection{Parallel and perpendicular magnetic field}
			
			So far we performed all tests with a magnetic field $45^\circ$ to the temperature gradient. To exclude the arbitrariness of this choice and to study in more detail the different implementations we carry out the tests also with two other setups:
			\begin{itemize}
				\item A magnetic field along the temperature gradient to check if the isotropic case can be recovered with the new code at sufficient accuracy.
				\item The other extreme case of a magnetic field perpendicular to the temperature gradient to see if the different implementations really recover total suppression of heat flux.
			\end{itemize}

			In the middle row of Fig. \ref{FIGconduction1all} we show the results for a parallel magnetic field again for both implementations. The fully anisotropic implementation recovers the analytic solution very well, however, with some noise. The amount of noise is about the same as with a the diagonal magnetic field. Since we have no difference to isotropic conduction in the case of a parallel magnetic field  we see, that the noise can not origin from computational instability due to a non positive definite transport matrix. In comparison, we find less scatter but the same expected offset in the isotropised run as before.
			
			In the bottom row we show similar plots for a magnetic field perpendicular to the temperature gradient. From our preconditions we expect no conduction in this  case, so the initial conditions should stay constant except for numerical noise.
			
			For the fully anisotropic derivation we find a rather stable solution. However, we encounter the regime, where the anisotropy is strong enough for heat to flow in the wrong direction. \nr{What we see here is not a numerical instability, since we eliminated the criterion of positive definiteness by using the advanced bi-CG solver, but rather an effect of the chosen discretisation.} We counter that behaviour and implemented the isotropised variation as a possible fix. 
			\nr{Please keep in mind that in these tests we did not switch on our satured heat flux limiter on purpose, to better see the effects of our discretisation on it's own. Besides that not being the case in \"real\" simulations, we do not expect to see this behaviour in simulations anyway, where other, more dominant, processes are included and immediately damp numerical instabilities. We will futher investigate how this error evolves with time in section \ref{timeevolutiontest}.}
						
			The isotropised approach can by construction not show a stable solution for this setup: The anisotropic part may be suppressed, but the isotropic part continues to work independently of the magnetic field. We find that while the conduction parallel to magnetic field lines is damped, we gain an increase of the perpendicular component. These violate the initial assumptions of our derivation for the sake of enforcing a physical heat flux. We have to consider different test setups to find out, which formulation gives better results and can be used for cosmological simulations.
			
			\begin{table}
				\centering
				\begin{tabular}{c||c|c}
					$\bmath{B}$ direction&$\Delta_{\mathrm{full ~ aniso}}$ [\%]&$\Delta_{\mathrm{isotropised}}$ [\%]\\
					\hline\hline
					No & \multicolumn{2}{c}{0.15} \\
					\hline
					$45^\circ$ & 0.37 & 0.84\\
					$0^\circ$ & 0.38 & 2.2\\
					$90^\circ$ & 2.2 & 4.1
				\end{tabular}
				\caption{Mean calculated relative errors of each particle's internal energy with respect to the analytic solution for the temperature step problem.}
				\label{TABsteperrors}
			\end{table}
			
			In table \ref{TABsteperrors} we present a more quantitative analysis of these results. For each particle inside the displayed range of $x \in [40, 60]$ cm we calculate the relative error in internal energy and collect the mean values for each displayed simulation. All shown errors lay at a percent level or below. Additionally, the isotropised version is about a factor two to ten worse than the fully anisotropic approach in all cases. For comparison we included the same calculation also for the reference run without magnetic field, were we reach a slightly smaller error value of about a factor two.
				
			For later simulation times the errors in the isotropised version rise slightly until the temperature profiles start to settle down to the isothermal convergence state, while the fully anisotropic version in general tends to converge stronger to the analytic result.
						
		\subsubsection{Random magnetic field}
			
			As last configuration, we check whether our implementation reproduces the common idea to approximate anisotropic conduction by an isotropic implementation damped by a factor of $1/3$. We imprint a random magnetic field onto our initial problems and fit the analytic solution with fitting parameter $\kappa$ to it. We find factors of about 0.33 at different times matching our expectations. The fully anisotropic approach usually shows slightly smaller values than the isotropised version, however the difference is very small. We emphasize that besides scatter, the shape of the analytic solution is reproduced very well. We conclude that unphysical errors like in the bottom left part of Fig. \ref{FIGconduction1all} will probably not arise in simulations with turbulent magnetic fields.

		\subsubsection{Time evolution}
			\label{timeevolutiontest}
			
			\nr{
			So far we only looked at a single snapshot of the internal energy's time evolution. In order to judge check that our conduction algorithm also behaves over long timescales and especially what happens with a perpendicular magnetic field we let the simulations run for a longer time and plot the result in figure \ref{FIGconduction1evolution}.
			
			\begin{figure*}
				\centering
				\includegraphics[width=\linewidth,rotate=90]{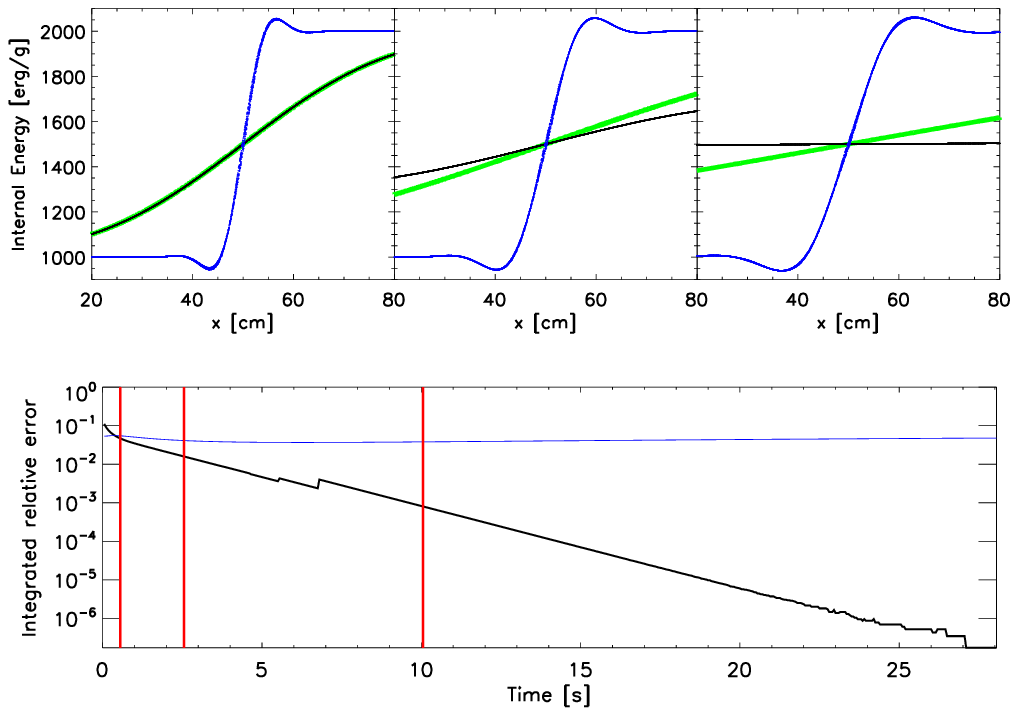}
				\caption{In this plot we show the time evolution of the temperature step test with the fully anisotropic formulation. The green line shows the analytic solution, in case of a diagonal magnetic field, integrated from minus until plus infinity, the black line presents the according simulation (with a finite box x-length of course, see text for further explanation) and the blue line depicts the simulation with a perpendicular magnetic field. The upper panel plots three snapshots in time which are marked by vertical lines in the lower panel, where the integrated relative error of both simulations compared to the solution are plotted.}
				\label{FIGconduction1evolution}
			\end{figure*}
			
			In the top layer we plot the analytic solution (green) and simulation (black) for the diagonal magnetic field and the simulation for the perpendicular magnetic field (blue) at three points in time, each further evolved than the plots we showed in the last subsections.
			
			The first thing to notice is, that while the shape of the black and green lines is always similar, the simulation seems to evolve much faster and settles for an isothermal state in the third plot while the analytic solution still harbours a significant temperature gradient. This is easily explained by the fact, that the analytic calculation actually does not match the simulation any more at this time. Recalling the derivation in appendix \ref{appendixtemperaturestep} we see that we integrated along the x-axis from $- \inf$ to $\inf$ in order to receive an easy result. However, the simulation box is of finite length in the x-direction. Namely it ranges between $x \in [0, 100]$. Up until now we plot only a subrange of this total width because of the same reason that the solution starts to become inaccurate at the boundaries. One can basically understand the difference as the solution harbouring an infinite reservoir of heat outside these boundaries, while the simulation has vacuum boundary conditions. In this plot we reached the time where we can see the influence of this reservoir become significant by pumping in more heat into the system from one side and providing a sink on the other one and therefore slowing down the convergence to an isothermal state in an accelerated manner. We elaborate briefly, why the integration boundaries can not be simply changed to account for that in appendix \ref{appendixtemperaturestepbounds}.
			
			As we can further see in this panel, while the \"bumps\" in the perpendicular configuration run formed quite quickly they present themselves as extremely stable: The magnitude of the error does not change at all over the whole evolution the only effect is that it broadens out a bit. The error is therefore not at all any kind of instability and it's evolution is strongly suppressed on it's own. Please note that the timescale on which we judge here is extremely long, since conduction becomes more and more ineffective when approaching the isothermal state.
			
			Finally the lower panel shows the time evolution of the integrated relative error of both simulations w.r.t. their analytic equivalent. We plot the integrated error because this resembles the conservation of energy. In addition we mark the three times at which the top panel plots were taken with vertical lines in this plot. The integrated error for the diagonal magnetic field stays constant over most of evolution after a small acclimatisation period which results from the discontinuity and a slightly asymmetry in sampling around it with SPH particles. For the perpendicular magnetic field the integrated error even declines exponentially. This perfectly displays how well our code is able to preserve the symmetry of the problem even though the particle setup is not based on a grid but on a glass file.			
			}

	\subsection{Smooth temperature distribution}

		Since the temperature step test contained an artificial discontinuity we test the code also with a similar setup but taking a smooth temperature distribution. Following \cite{Cleary1999} we take a sinusoidal temperature distribution at $t=0$.
		At first, we derive the analytic solution for the initial conditions:
		\begin{equation}
			u_0(x') = u_0 \cdot \sin \lp k x \rp
		\end{equation}
		with a generic wavenumber $k$. The result is:
		\begin{equation}
			u(t, x) = u_0 \sin \lp kx \rp e^{- \alpha k^2 t} .
			\label{EQgeneralsinusoidalend}
		\end{equation}
		
		Assuming periodic boundary conditions we need to add an initial offset to prevent negative energies:
		\begin{equation}
			u(t,x) = u_1 + u_0 \sin \lp 2 \pi \frac{x}{L} \rp e^{- 4 \pi^2 \alpha t / L^2} .
			\label{EQsinusoidalperiodicfinal}
		\end{equation}
		
		We chose the arbitrary values of $u_1 = 1500$ erg/g and $u_0 = 1000$ erg/g. Including a magnetic field we expect a reduced conduction with coefficient \mbox{$\kappa' = \kappa \cdot \cos^2\angle\lp\bmath{B}, \bmath{\nabla} T\rp$.}

		\begin{figure*}
			\centering
			\includegraphics[width=\linewidth]{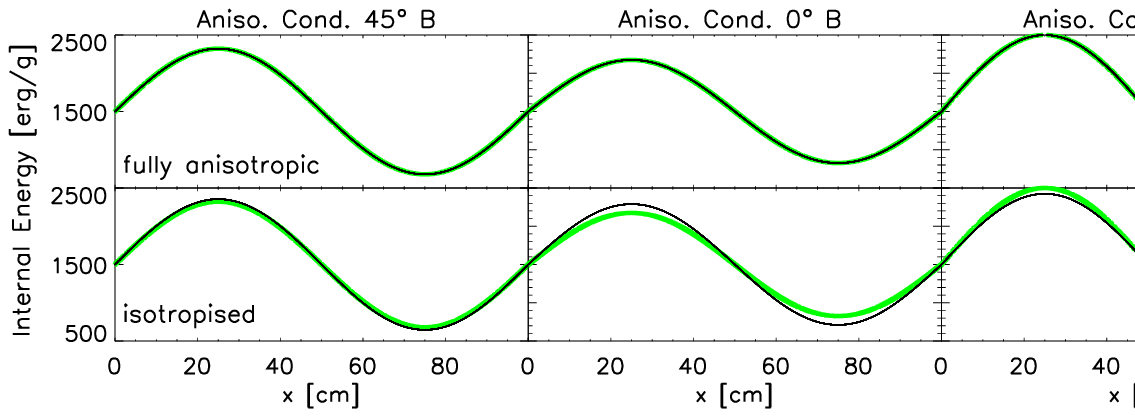}
			\caption{Testing a wave like temperature distribution as initial conditions using both the full anisotropic approach (upper row) and the isotropised version (lower row) with three different magnetic field configurations: $45^\circ$ to the temperature gradient (left column), parallel (middle column) and perpendicular (right column), all at the same simulation time. Basically, the results are the same as for the temperature step test. However, we see less scatter in our plots for this smooth setup. Even the error which we saw for the perpendicular magnetic field earlier does not appear in this test.}
			\label{FIGconduction2}
		\end{figure*}
		
		We perform this test with both implementations of anisotropic conduction and three magnetic field configurations with $0^\circ$, $45^\circ$ and $90^\circ$ to the $x$-axis. The results are shown in Fig. \ref{FIGconduction2}.
		
		Basically, we find a similar behaviour as for the temperature step problem: The isotropised impelmentation shows always an offset from the analytic solution (weaker conduction parallel and stronger conduction perpendicular to the magnetic field), while the fully anisotropic implementation reproduces the solution very well. We emphasise two main differences to our previous results:
		
		The amount of scatter for the fully anisotropic implementation is similar to what we get for the isotropised run. Since there is no strong discontinuity in this setup the amount of scatter is way lower than for the temperature step test.
		
		Furthermore, we do not get any numerical artifacts in our results and even full suppression of conduction with a perpendicular magnetic field for the fully anisotropic implementation. Therefore, this approach produces the best results as long as there are no sudden temperature jumps.
		
	\subsection{Hot gas sphere}
	
		Next, we test how the code behaves for a more complex scenario. Similar to the second test from \cite{Jubelgas2004} we set-up a sphere of hot gas. We use spherical symmetric initial conditions for the internal energy in the form of
		\begin{equation}
			u_0(r) = u_0 e^{-\beta r} .
			\label{EQinitialsphere}
		\end{equation}
		For this test case we only show a qualitative comparison of the different runs, to see if the anisotropy is reproduced well.

		\begin{figure*}
			\centering
			\includegraphics[width=\linewidth]{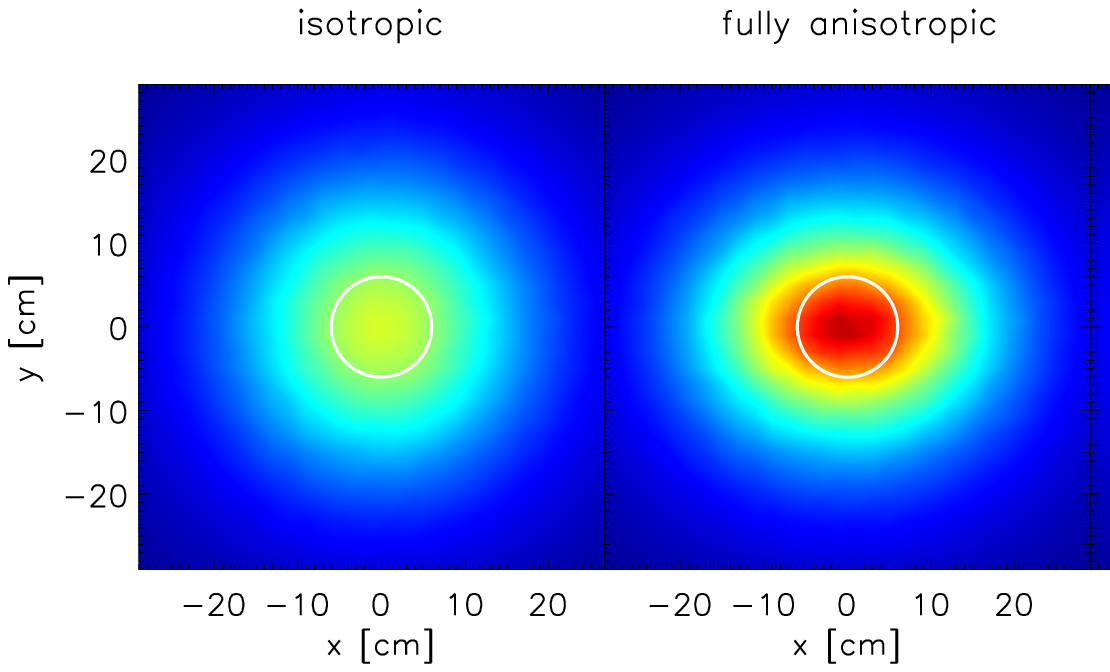}
			\caption{Testing a spherical symmetric temperature distribution with a radial gradient. The left plot shows the reference run with isotropic conduction, while the middle and right plot show the fully anisotropic and isotropised implementations both with a magnetic field in $x$ direction, again all at the same time. One can see, that the overall conduction increases with isotropy of the implementation. The much stronger anisotropic approach in the middle plot can be seen through higher ellipticity of isotherms (illustrated by white circles).}
			\label{FIGconduction3}
		\end{figure*}
		
		In Fig. \ref{FIGconduction3} we show our test results using isotropic conduction and both anisotropic approaches for a magnetic field in $x$ direction. The comparison shows well the different overall effect of the three implementations. The more anisotropy the approach contains the lower the temperature decline in the inner region. Additionally, we see the stronger anisotropy in the middle panel compared to the right one through the ellipticity of the resulting profile.
		
		In total, this agrees with our previous findings. Again there are no strange artifacts visible in any of the runs. We conclude that the fully anisotropic approach should be fairly unproblematic to use while it gives us more exact results according to the properties we formulated at the beginning. Therefore, we consider only this formulation.
		
	\subsection{Temperature step with perpendicular Conduction}
		\label{tempstepperp}
		
\begin{figure}
			\centering
			\includegraphics[width=\linewidth]{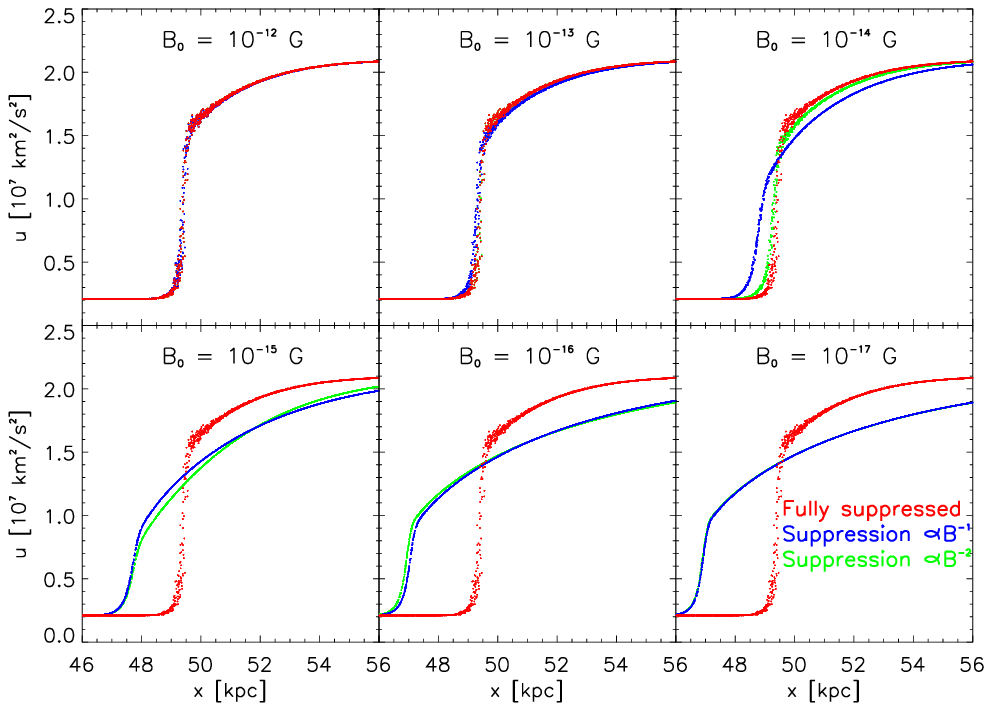}
			\caption{Temperature step problem for different magnetic field strengths and treatments of perpendicular conduction. All plots are made at the same simulation time. One can clearly identify magnetic field strengths where the relation between linear and quadratic suppression flips concerning which prescription results in higher net conduction.}
			\label{FIGconduction4}
		\end{figure}

		Finally, we again set up a temperature step problem but now we investigate the behaviour of the suppression mechanism described in section \ref{anisotropicthermalconduction}. We use typical values for temperature and density as they are found in hot regions of galaxy clusters and use a homogeneous magnetic field of the form $\bmath{B} = \{B_0, B_0, B_0\}$ with \mbox{$B_0 \in \left[10^{-12} \mathrm{G}, 10^{-17} \mathrm{G} \right]$ }. The results are shown in Fig. \ref{FIGconduction4}.
		
		At the beginning we set up the temperature step at $x = 50$ kpc. The generally expected behaviour is, that the discontinuity propagates to the low temperature regime while the two levels close in on the mean temperature. We have run the different set-ups with either totally suppressed conduction perpendicular to the magnetic field and both phenomenologically motivated scaling relations presented in section \ref{anisotropicthermalconduction}.
		
\begin{figure}
			\centering
			\includegraphics[angle=270,width=\linewidth]{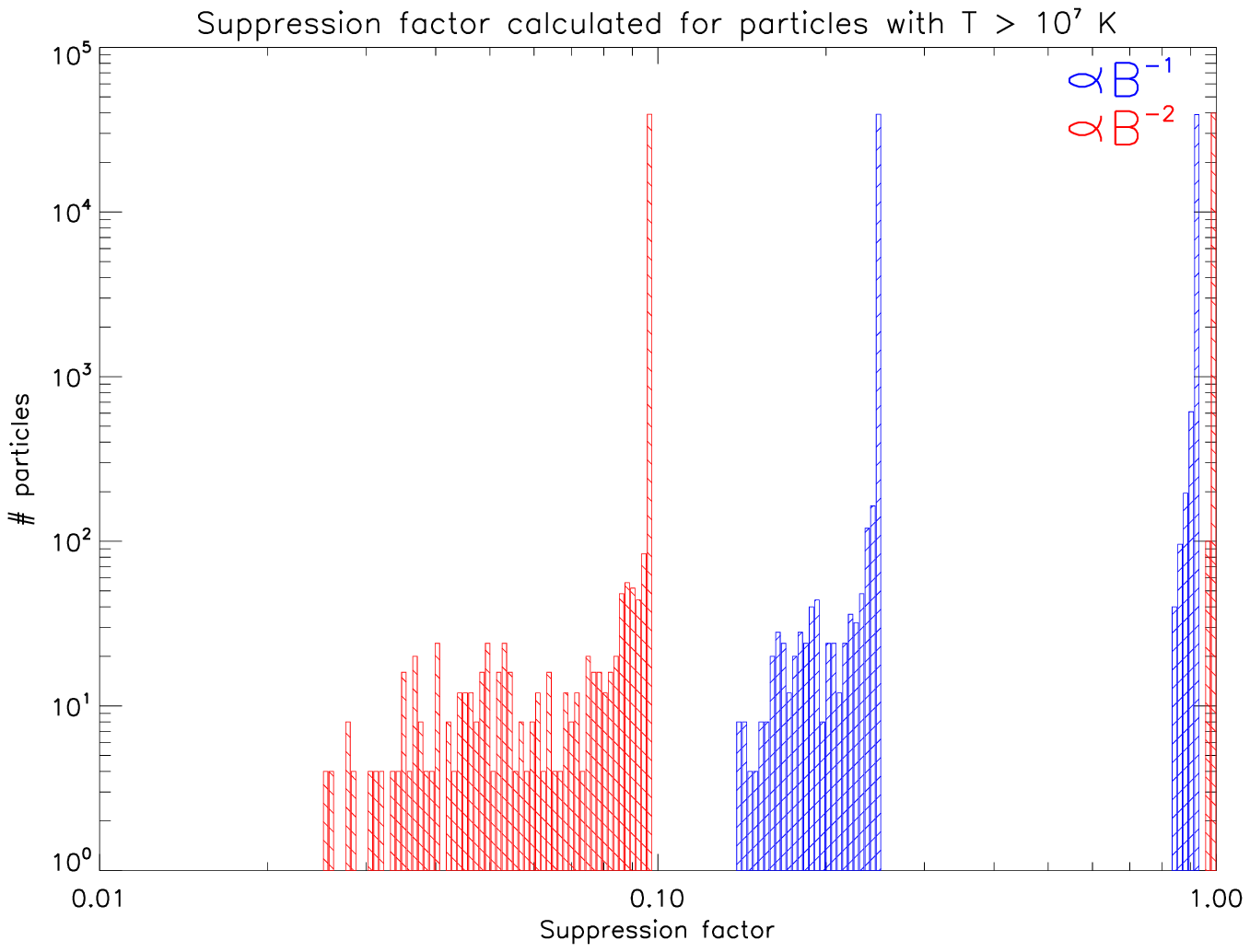} Initial conditions
			\caption{Number of particles per suppression factor bin for the initial conditions with $B_0 = 10^{-15} \mathrm{G}$. Particles at the higher plateau get a stronger suppression for the quadratic formalism while particles at the lower plateau show the opposite behaviour.}
			\label{FIGconduction4hist}
		\end{figure}

		We can identify the following different behaviours when varying the magnetic field strength:
		\begin{itemize}
			\item $B_0 \ge 10^{-13} \mathrm{G}$: The magnetic field is strong enough to fully suppress perpendicular conduction no matter which prescription we use.
			\item $B_0 \sim 10^{-14} \mathrm{G}$: The linear scaling relation results in an increased net conduction while the quadratic scaling still suppresses perpendicular conduction strongly.
			\item $B_0 \sim 10^{-15} \mathrm{G}$: Both prescriptions allow a certain amount of perpendicular conduction however there is no clear relation between both. The denominator of the suppression factor on the higher energy level is larger than one which results in a stronger suppression when the factor is squared. However, it is smaller than one for the low energy level. This is illustrated by Fig. \ref{FIGconduction4hist}.
			\item $B_0 \sim 10^{-16} \mathrm{G}$: The relation between linear and quadratic scaling has fully flipped: While both allow for a lot of perpendicular conduction now we get more net conduction with the quadratic formula
			\item $B_0 \le 10^{-17} \mathrm{G}$: The magnetic field is so weak that it can not suppress perpendicular conduction any more with either of the discussed scaling relations.
		\end{itemize}
		
		In total, we see that a proper treatment of perpendicular conduction is important mostly for very small magnetic field strengths. We can not judge from this test which prescription is the better, however, it is important to include a prescription if small magnetic fields require proper treatment. Additionally, we note that even if we take into account only hot gas, the suppression is still also dependent on density, which means that also particles with stronger magnetic fields can require this proper description.
		
	\subsection{Summary of test results}
	
		After all tests we come to the following conclusions:
		
		\begin{itemize}
			\item The isotropised formulation for anisotropic conduction ensures that the solving algorithm is stable and does not lead to non-physical heat conduction. However, it violates the prerequisites we used to derive an anisotropic formulation.
			\item We find that the fully anisotropic formulation behaves sufficiently well for adequately smooth temperature distributions. \nr{Since the degree of instability should be small in comparison to hydrodynamical effects and is further suppressed by the application of a heat flux limiter, we only apply this formulation in our cosmological simulations. We made sure that the instability does not grow over time on it's own and is furthermore not a numerical one in sense of the solver we apply to the discretisation of the conduction equation.}
			\item We have briefly investigated the effects of different scalings for perpendicular suppression and further inquire their behaviour in simulations of galaxy clusters.
		\end{itemize}

\section{Application to galaxy clusters}
	\label{applicationgeneral}
	
	\begin{figure*}
		\centering
		\includegraphics[width=\linewidth]{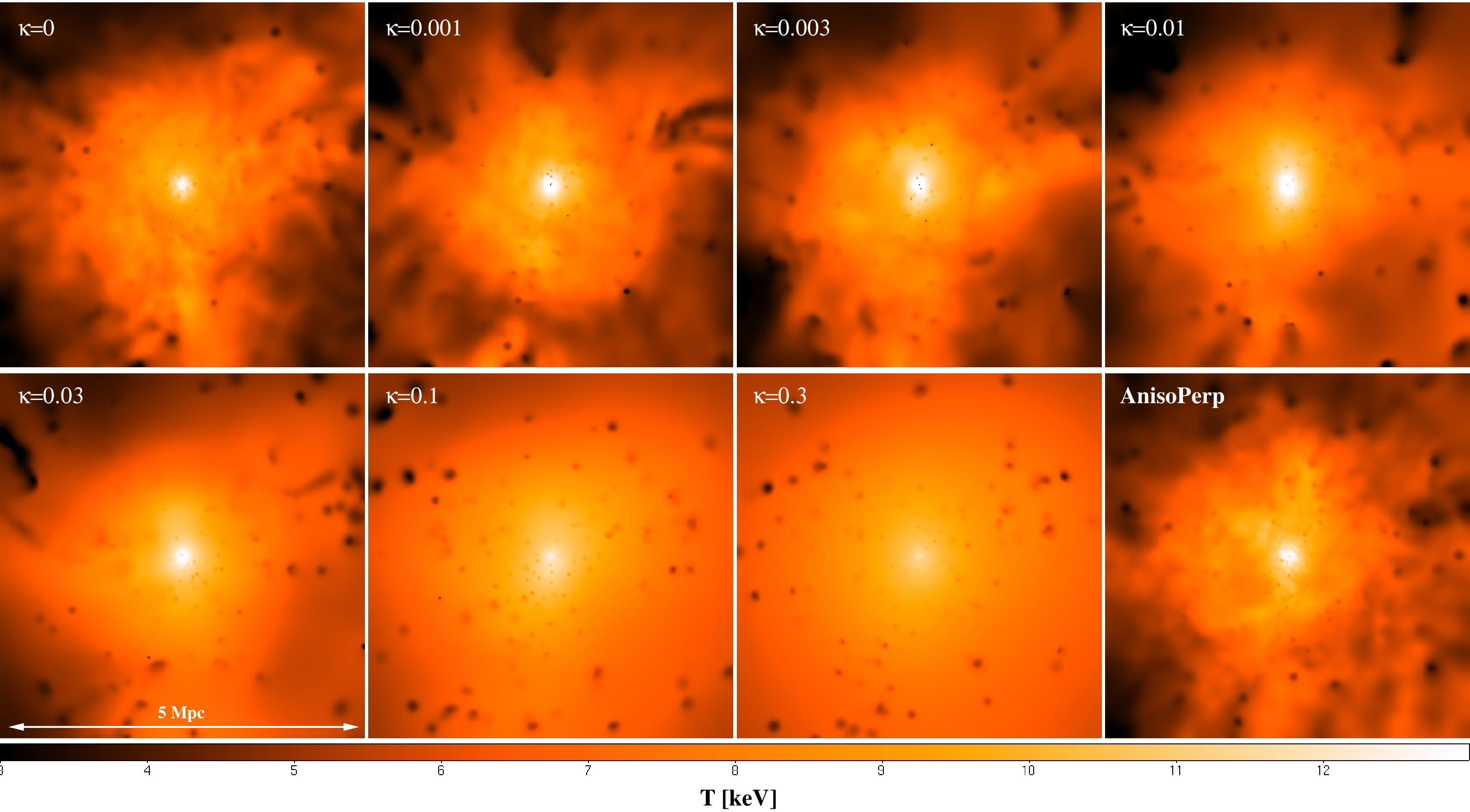}
		\caption{Shown are the mass weighted temperature maps (5 Mpc $\times$ 5 Mpc) of the relaxed cluster \mbox{\it g5699754} at z=0. The upper left panel shows the simulation without any thermal conduction. The other maps (from upper left top lower right) show the sequence for isotropic thermal conduction when changing the supression factor  as indicated in the maps. The lower right one shows the run with anisotropic thermal conduction where the perpendicular term is evaluated proportional to the magnetic field strength.}
		\label{FIGmapsg5699754}
	\end{figure*}
	
	In this section we present zoomed in re-simulations of massive COMA-like galaxy clusters selected from large Gpc sized cosmological boxes, where the parameters for a  $\Lambda$CDM cosmology with $\Omega_m = 0.24$, $\Omega_b = 0.04$, $\Omega_\Lambda = 0.76$ and $h = 0.72$.
    We select five clusters from the original set of simulations presented in \cite{Bonafede2011} to study the effect of thermal conduction within galaxy clusters.
    These galaxy clusters have virial masses\footnote{Here we define virial properties based on the averaged density as predicted by the top-hat spherical collapse model, corresponding to 95 times the critical density for our chosen cosmology.} of  $M_\mathrm{vir} > 10^{15} M_\odot /h$ corresponding to virial radii of  typically $r_\mathrm{vir} \sim 2.5 \; \mathrm{Mpc} / h$.
    Calculated virial properties for all runs are listed in Table \ref{TABcluster}.
    More details about the selection of the galaxy clusters and the generation of the initial conditions can be found in \cite{Bonafede2011}.
	
\subsection{The effects of different conduction prescriptions}

    At first, we select one isolated and relaxed looking galaxy cluster (\mbox{\it g5699764}) to perform several simulations testing different settings for the implementation of thermal conduction.
    Fig. \ref{FIGmapsg5699754} displays projected temperature maps of 5 Mpc wide and thick slices through the cluster, demonstrating the effect of thermal conduction on the temperature structure.
    The upper left panel shows the reference run without any thermal conduction.
    Then, from left to right and top to bottom the suppression factor is reduced (i.e. the conduction efficiency is increased) for the case of isotropic heat conduction.
    The last panel bottom right shows the result for anisotropic heat conduction, where we include the linearly scaling perpendicular suppression factor as displayed before (see section \ref{anisotropicthermalconduction}).
    
    As shown already in \cite{Dolag2004}, where similar simulations have been carried out with an earlier version of the implementation of isotropic heat conduction, we see that in such massive (and therefore hot) galaxy clusters, isotropic conduction has a strong effect on the temperature distribution.
    With less suppression of thermal conduction, more heat gets transported from the central part of the cluster to the outskirts and even more dramatically visible, local temperature fluctuations get smoothed out.
    In contrast, the simulation with anisotropic heat conduction shows only a very mild smoothing of the temperature fluctuations compared to the control run.
    
    \begin{figure}
    	\centering
    	\includegraphics[width=\linewidth]{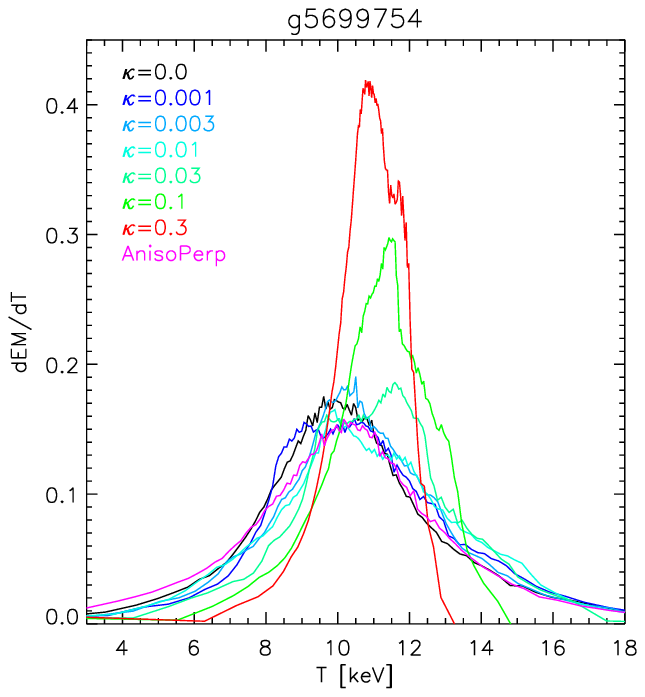}
    	\caption{Shown are the emissivity distribution of the ICM within the virial radius as function of the temperature for different treatments of the thermal conduction. Good to see the trend to shrink the distribution around the mean temperature with increasing level of the coefficient.}
    	\label{FIGemissivity1}
   	\end{figure}
   	
   	This can also be seen in Fig. \ref{FIGemissivity1}, where we show the emissivity  distributions for all the isotropic and one anisotropic run.
   	The larger the isotropic conduction coefficient the more the distribution is taylored around the mean temperature, (e.g. the cluster gets more isothermal), while the peak increases and shifts to slightly
   	higher temperatures.

   	\begin{figure}
   		\centering
   		\includegraphics[width=\linewidth]{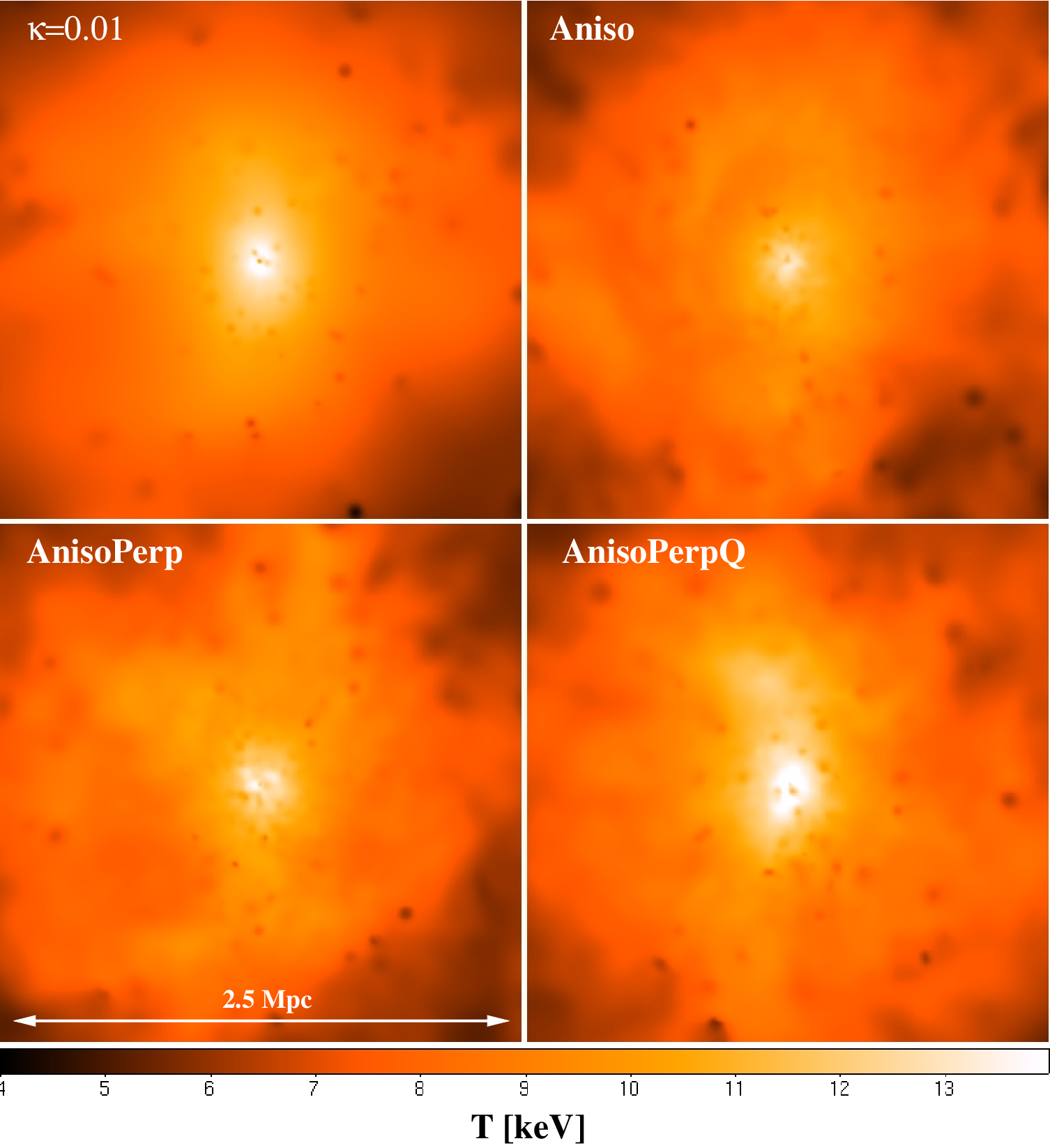}
   		\caption{Shown are the mass weighted temperature maps (2.5 Mpc $\times$ 2.5 Mpc) of the inner part of the relaxed cluster at $z=0$. The upper left panel shows the simulation with isotropic conduction for $\kappa=0.01$. The other three maps show the runs with anisotropic thermal conduction for different treatment of the perpendicular case (see section- \ref{anisotropicthermalconduction}).}
   		\label{FIGmapsg5699754zoom}
   	\end{figure}
   	
   	To investigate the effect of details in the different treatment of perpendicular conduction for the anisotropic heat conduction, Fig. \ref{FIGmapsg5699754zoom} shows temperature maps zooming onto the central 2.5 Mpc of our test cluster.
   	Here we compare the isotropic thermal conduction with a suppression factor of $\kappa = 0.01$ with three anisotropic runs with different treatment of the perpendicular component: fully suppressed (\mbox{\it Aniso}), linear (\mbox{\it AnisoPerp}) and quadratic (\mbox{\it AnisoPerpQ}) proportionality to the magnetic field strength.
   	It is clearly visible that the detailed choice of treatment of the perpendicular component has a quite significant effect on the outcome.
   	Still, none of the anisotropic runs show such a strong smoothing of local temperature fluctuations as the isotropic conduction simulation, even if we allow for conduction to become rather isotropic for weak magnetic fields.
   	It also makes a notable difference if we use the linear or the quadratic formula to calculate the perpendicular suppression factor.
   	
   	\begin{figure}
   		\centering
   		\includegraphics[width=\linewidth]{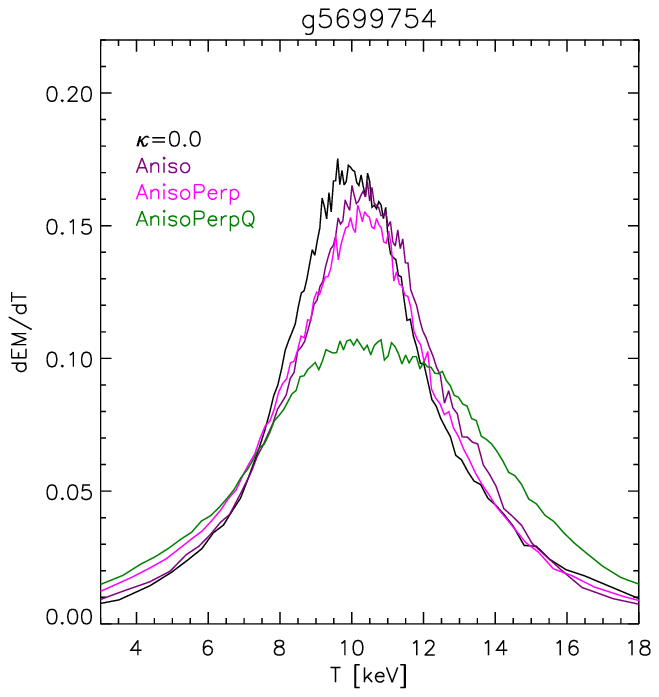}
   		\caption{Shown are the emissivity distribution of the ICM within the virial radius as function of the temperature for runs with anisotropic thermal conduction for the different treatments of the perpendicular case (see section \ref{anisotropicthermalconduction}).}
 		\label{FIGemissivity2}
	\end{figure}
	
	This gets again more clear when looking at the emissivity distributions for the different anisotropic runs shown in Fig. \ref{FIGemissivity2}.
	While including a perpendicular suppression coefficient proportional to $B^{-1}$ shrinks the distribution a bit, since it contains overall more conduction, we see clearly a different picture for the case proportional to $B^{-2}$.
	We find that this prescription suppresses conduction perpendicular stronger for most particles which results in less conduction compared	to the linear case.
	Therefore, we see the emissivity distribution broadening
	again, even beyond the case with zero thermal conduction.
	
	\begin{figure}
		\centering
		\includegraphics[angle=270,width=\linewidth]{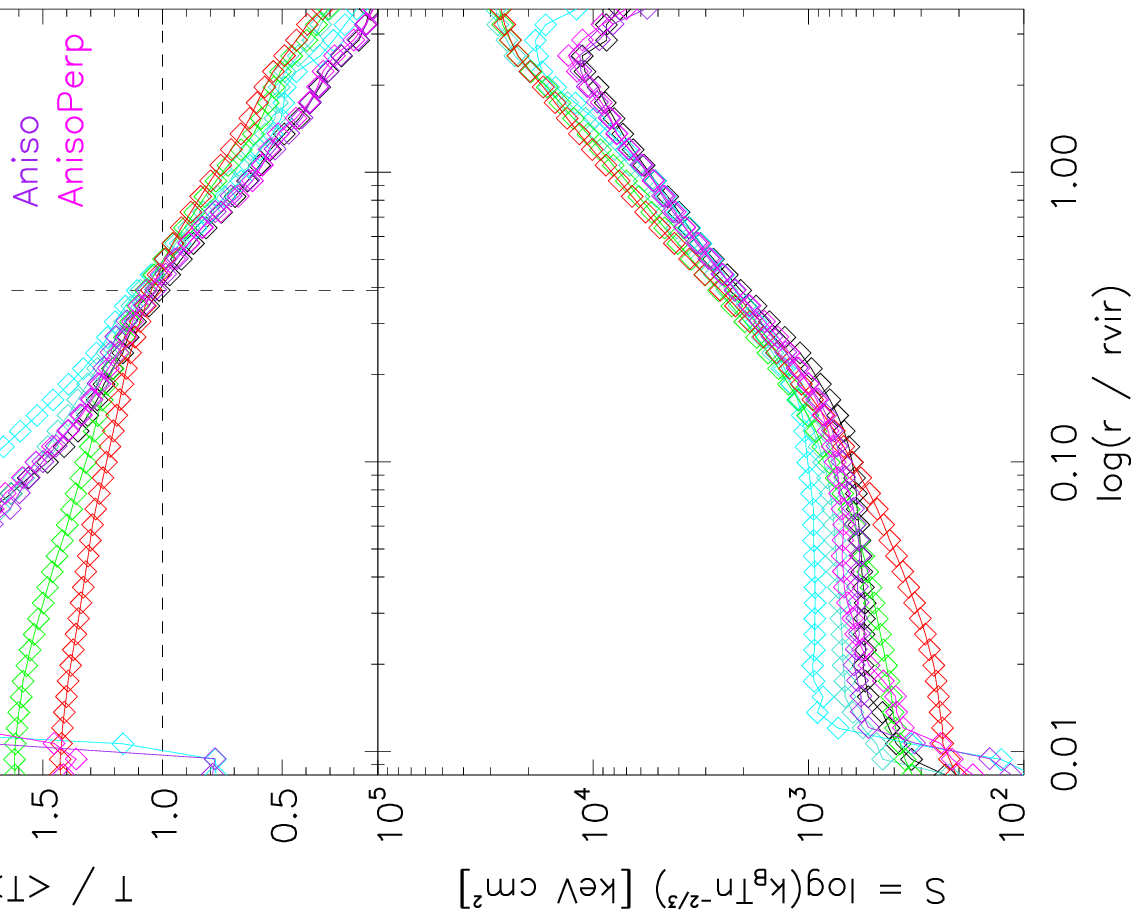}
		\caption{Shown are radial temperature (upper panel) and entropy (lower panel) profiles for \mbox{\it g5699764} with several different conduction settings at $z=0$. The temperature profiles are normalized by the mean temperature within the virial radius for each respective run.}
		\label{FIGradialtempprofiles}
	\end{figure}
	
	A more quantitative analysis is shown in Fig. \ref{FIGradialtempprofiles}, where the scaled, radial temperature profiles are presented in the upper panel.
	Here it can be clearly seen that the stronger we choose the isotropic conduction coefficient, the more internal energy is transported outwards beyond the virial radius.
	In agreement with previous studies in \citep{Dolag2004}, isotropic thermal conduction at a level of 1/3 of the Spitzer value already leads to an isothermal temperature distribution in the inner part of the galaxy cluster.
	Regarding anisotropic conduction we include two runs, the one with full suppression perpendicular to magnetic field lines as well as the one using the linearly scaling suppression factor.
	While the totally suppressed run resembles almost zero level of isotropic thermal conduction, the temperature profile of the latter one lies somewhere between $\kappa \approx 0.01$ and $\kappa \approx 0.03$ for isotropic thermal conduction.
	The entropy profiles in the lower panel are in general more difficult to interpret.
	Here, already the reference run builds up a significant entropy core (in line with what was reported for the non ideal MHD simulations in \cite{Bonafede2011}).
	Their shape is rather similar for all runs and due to combined effects including the different implementations of thermal conduction the trends are not easy to interpret and seem to depend on the local dynamical structures in the core of the galaxy cluster.
	
	\begin{figure}
		\centering \includegraphics[width=\linewidth]{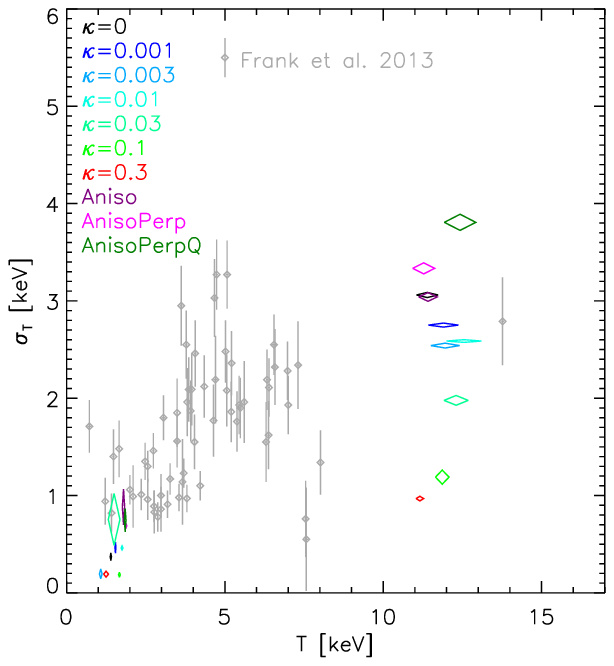}
		\caption{Comparing the temperature fluctuations within $R_{2500}$ as inferred from observations by \protect\cite{Frank2013} with the one predicted for the simulated relaxed cluster with the different treatment of the thermal conduction (as labelled).}
		\label{FIGcomp1}
	\end{figure}
	
	Finally, we can compare the simulated cluster to a sample of observations
	with \textit{XMM-Newton} presented by \cite{Frank2013}, where they measured the width of the temperature fluctuations within the central part (e.g. within $R_{2500}$) of a sample of galaxy clusters.
	Fig. \ref{FIGcomp1} shows these observational data points over plotted with results for our different implementations of thermal conduction within our simulations.
	Please note that here we not only use the central galaxy cluster but also make use of a smaller galaxy cluster present within our simulation, which has a temperature	of roughly 1 keV.
	For the case of isothermal conduction we see that the high isotropic conduction coefficients (e.g. $\kappa=0.3$ and $\kappa=0.1$) produce results which are below the observed temperature fluctuations for the high temperature system, similar to the findings in \citet{Rasia2014}.
	For the low temperatures all implementations are consistent with the observations.
	In contrast, the anisotropic runs are matching with the simulations without	thermal conduction.
	Interestingly, the simulation with the quadratic dependency	of the suppression factors shows the largest temperature fluctuations, in line
	with the broader temperature distribution shown before.
	
	\begin{figure*}
		\centering
		\includegraphics[angle=270,width=\linewidth]{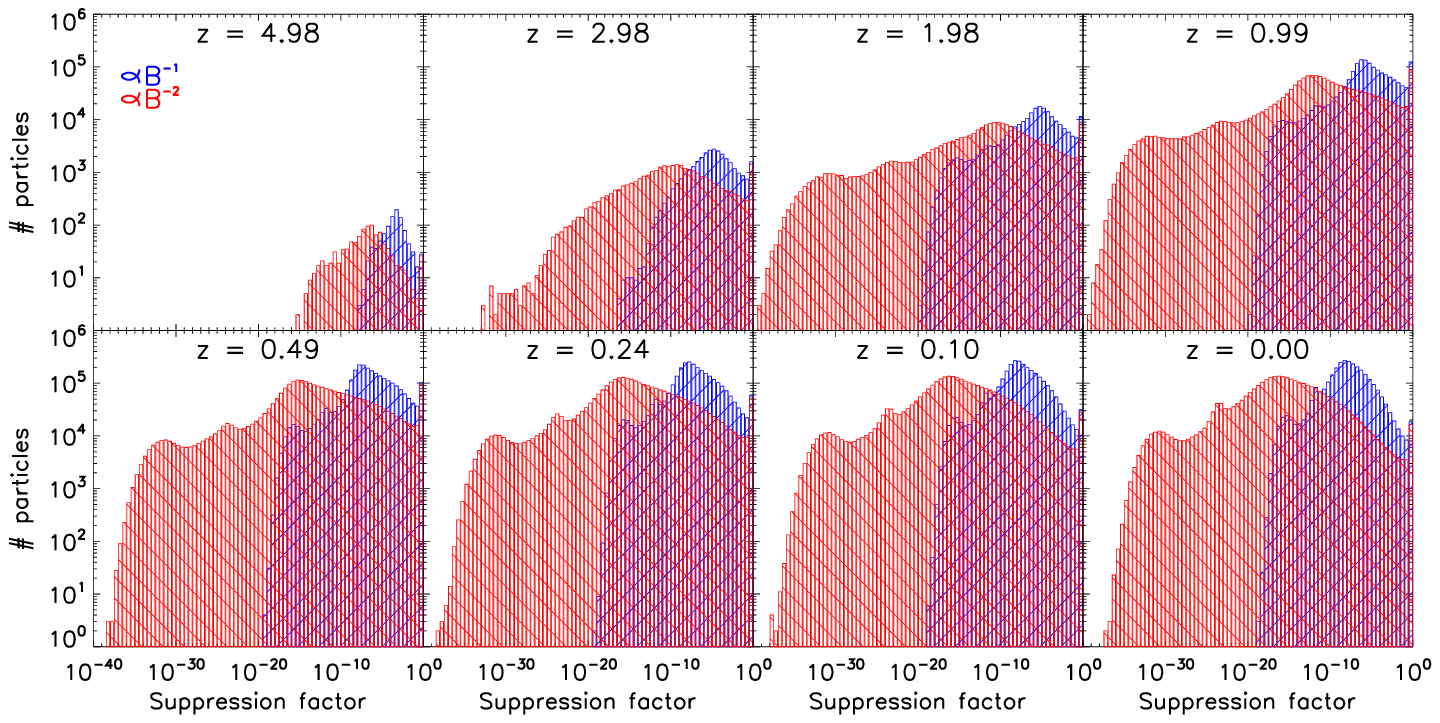}
		\caption{This plot shows distribution of linear and quadratic suppression factors calculated for all particles of the run AnisoPerp with temperatures higher than $10^7$ K for several redshifts. (see section \ref{anisotropicthermalconduction}).}
		\label{FIGsupfac}
	\end{figure*}
	
	To enforce the idea of a proper treatment of perpendicular conduction we display the range of suppression factors for all hot particles in	one of our simulations in Fig. \ref{FIGsupfac} at several
	redshifts.
	Since conduction scales strongly with the temperature of the plasma, we take into account only the most important contributors to thermal conduction, selecting particles within the hot atmosphere of groups and clusters by requesting their temperatures $T$ to exceed $10^7$ K.
	As the typical formation time of clusters and groups is around $z = 1$, the amount of particles within this hot gas phase increases significantly until the redshift approaches $z = 1$.
	While either suppression formulation results in fairly low suppression factors for the bulk of particles, there are significant differences in the amount of particles which have moderate	suppression factors up to the regime of almost unsuppressed conduction.
	Furthermore, it seems that the quadratic formula produces in general lower factors and therefore less net conduction than the linear one. As we have already seen in our tests in section \ref{tempstepperp} the two formulations show opposite behaviour in different regimes.
	
	To further investigate the effects of the isotropic and anisotropic treatment of thermal conduction we select the same four, Coma like galaxy clusters ({\it g0272097}, {\it g1657050}, {\it g4606589}	and {\it g6802296}) as in \citet{Bonafede2011} and simulate them with zero thermal conduction, isotropic conduction at a level of $\kappa=0.3$ and anisotropic thermal conduction using the linear scaling for the perpendicular case.
	Table \ref{TABcluster} lists the general properties of the resulting galaxy clusters.
	
	\begin{table}
		\centering
		\begin{tabular}{l|l|l|l|l|l|l}
			Cluster & Bon. & Cond. $\kappa$ & $r_\mathrm{vir}$ & $m_\mathrm{vir}$ & $f_\mathrm{col}$\\
			\hline
			\hline
			g5699754 & D17 & $0$ & 2.568 & 1.739 & 0.184\\
			&     & $0.001$ & 2.566 & 1.736 & 0.183\\
			&     & $0.003$ & 2.569 & 1.742 & 0.175\\
			&     & $0.01$ & 2.569 & 1.742 & 0.175\\
			&     & $0.03$ & 2.566 & 1.735 & 0.181\\
			&     & $0.1$ & 2.569 & 1.741 & 0.188\\
			&     & $0.3$ & 2.564 & 1.732 & 0.191\\
			&     & Aniso & 2.553 & 1.710 & 0.197\\
			&     & AnisoPerp & 2.562 & 1.727 & 0.188\\
			&     & AnisoPerpQ & 2.553 & 1.709 & 0.183\\
			\hline
			g0272097 & D2 & $0$ & 2.375 & 1.3989 & 0.179\\
			&    & $0.3$ & 2.378 & 1.393 & 0.185\\
			&    & AnisoPerp & 2.310 & 1.280 & 0.267\\
			\hline
			g1657050 & D5 & $0$ & 2.397 & 1.427 & 0.185\\
			&    & $0.3$ & 2.387 & 1.410 & 0.191\\
			&    & AnisoPerp & 2.380 & 1.398 & 0.193\\
			\hline
			g4606589 & D13 & $0$ & 2.145 & 1.025 & 0.174\\
			&     & $0.3$ & 2.153 & 1.037 & 0.191\\
			&     & AnisoPerp & 2.154 & 1.038 & 0.180\\
			\hline
			g6802296 & D20 & $0$ & 2.062 & 0.909 & 0.172\\
			&     & $0.3$ & 2.054 & 0.899 & 0.204\\
			&     & AnisoPerp & 0.202 & 8.566 & 0.211
		\end{tabular}
		\caption{Calculated virial masses [$10^{15} \; M_\odot$/h], virial radii [Mpc/h] and mass fraction of collapsed baryons (stars + gas with $T < 3 \cdot 10^4 \; \mathrm{K}$) for all runs. Different conduction settings do not alter the global halo properties significantly. The fraction of collapsed baryons seems to grow slightly with increasing net conduction. For further cross reference our initial conditions with the numbering used by \protect\cite{Bonafede2011}.}
		\label{TABcluster}
	\end{table}
	
	While the virial properties of the halo are basically unchanged, the amount of condensated baryons in form of stars and cold gas changes with the treatment of thermal conduction.
	This fraction slightly grows with increasing net conduction, similar to previous findings \citep{Dolag2004}, again indicating that thermal conduction alone is not able to prevent cooling in the centers of cluster.
	Although, due to the inclusion of magnetic fields, the fraction of condensed baryons is smaller than in previous numerical studies, it is still larger than previous observations \citep{Balogh2001,Lin2003,Andreon2010}.
	However, more recent observational studies by \cite{Kravtsov2014} indicate a significantly larger amount of stars in the central galaxies of clusters than previously thought.
	Ultimately, including anisotropic thermal conduction seems not to change the amount of cold baryons in the center of simulated galaxy clusters significantly.
	
	\begin{figure*}
		\centering
		\includegraphics[width=\linewidth]{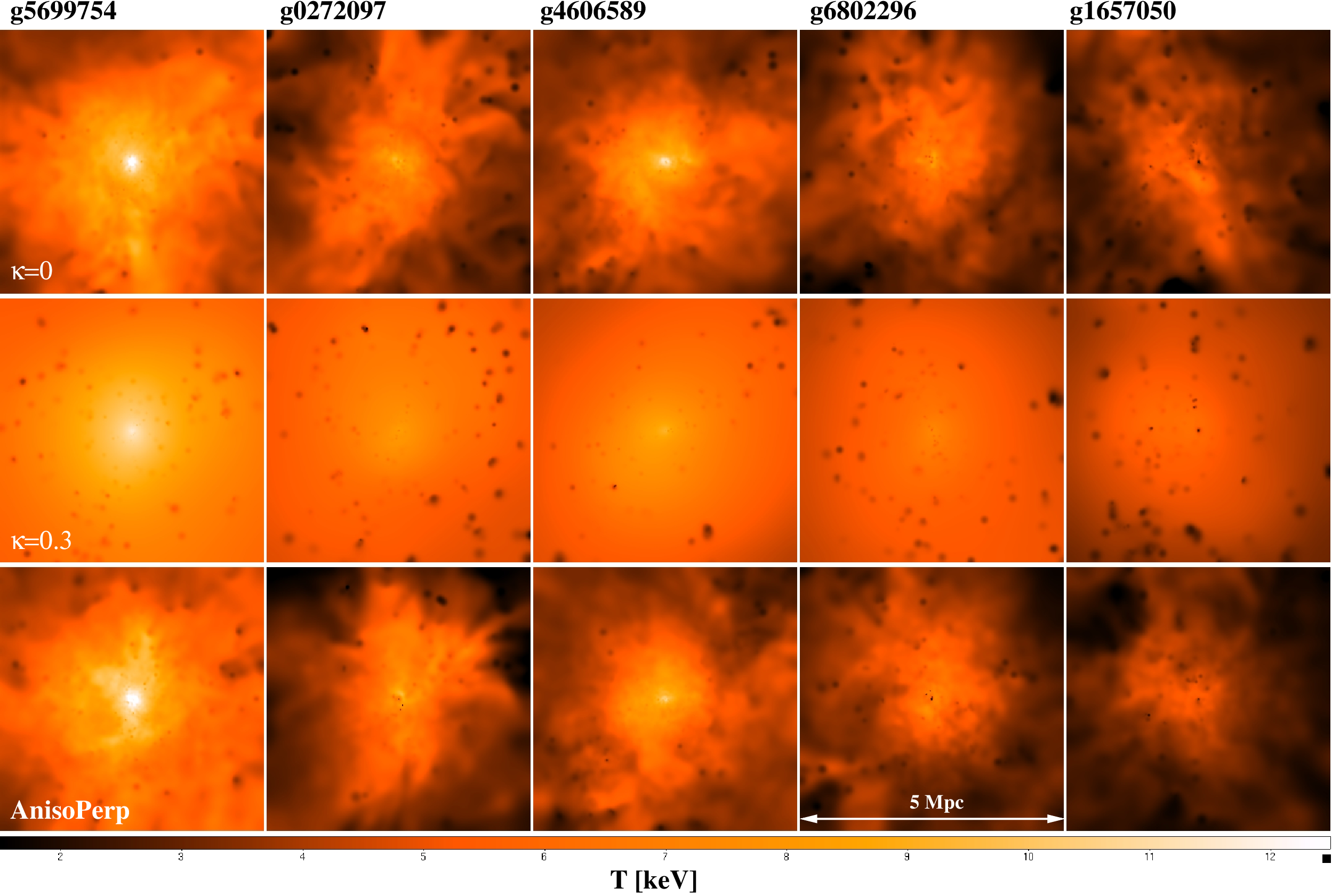}
		\caption{Shown are the mass weighted temperature maps (5 Mpc $\times$ 5 Mpc) of the five simulated clusters at z=0. The upper left row shows the simulation without thermal conduction. The middle maps show the simulations with isotropic thermal conduction for	$\kappa=0.3$ and the lower row shows the runs with anisotropic thermal conduction, where the perpendicular term is evaluated with the linear scaling.}
		\label{FIGmapsmoreics}
	\end{figure*}
	
	\begin{figure*}
		\centering
		\includegraphics[width=\linewidth]{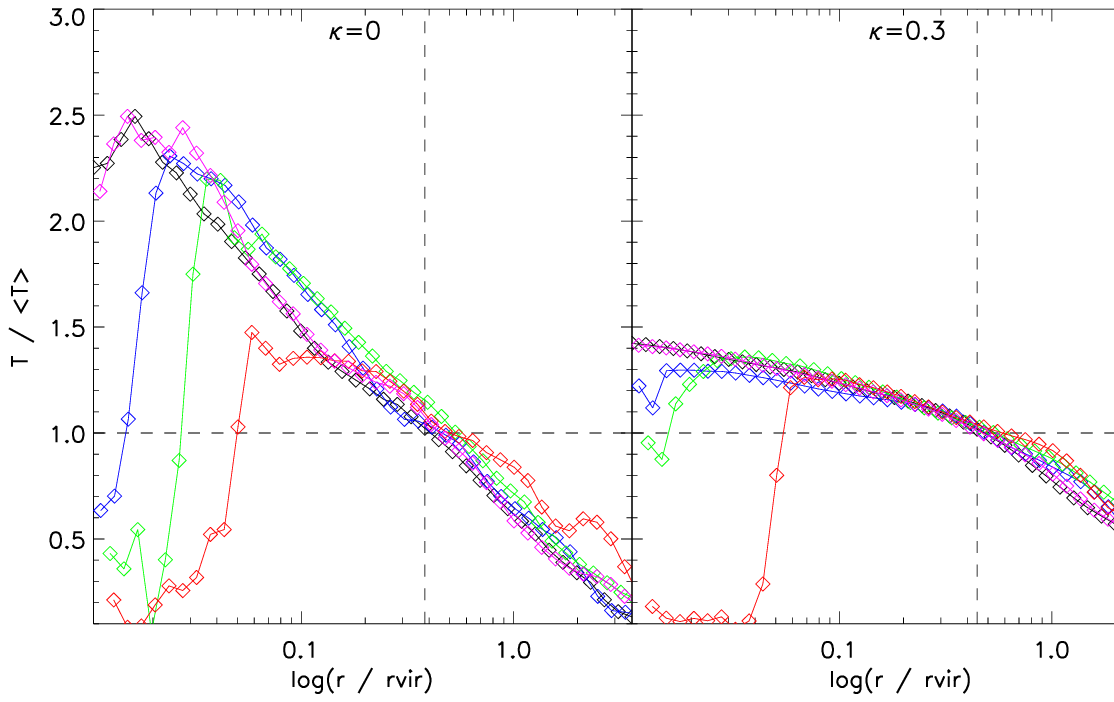}
		\caption{Radial temperature profiles for all five clusters as in
		Fig. \ref{FIGmapsmoreics}. Dashed lines indicate the first crossing of T = $\mathrm{<T>}$.}
		\label{FIGradialtempprofilesall}
	\end{figure*}
	
	The respective temperature maps for all five clusters with the three settings for thermal conduction are shown in Fig. \ref{FIGmapsmoreics}.
	The four additional clusters show a a very similar behaviour as we saw before in the relaxed one.
	Temperature is transported outwards with the isotropic conduction using $\kappa=0.3$, while substructures are strongly smoothed out, where as the run with anisotropic conduction shows only mild smoothing of temperature fluctuations.
	One interesting aspect gets	clearly visible in Fig. \ref{FIGradialtempprofilesall}, where we present the corresponding radial temperature profiles for five clusters for the three different runs.
	Again, we see that isotropic conduction leads to a significant flattening of the temperature profile embedding a cold core with varying size and moderate temperature.
	The simulations without thermal conduction show a rising temperature profile towards the center with a much larger drop of temperature within the central core.
	The simulations with anisotropic thermal conduction where we used the linear scaling for the perpendicular component shows a more bimodal temperature profile.
	Some clusters show very similar temperature profiles compared to the simulations without any	thermal conduction, some have a very pronounced cold core.
	The sample is much to small to draw robust conclusions but this indicates that in the case of anisotropic thermal conduction the amount of heat transport is strongly varying with the current dynamical state of the cluster and therefore might contribute to the observed bimodality of cool core and non cool core clusters.
	The temperature for all runs drops lower than the specific mean temperature of gas inside the virial radius at about 40 to 45 per cent of the virial radius.

	\begin{figure}
		\centering
		\includegraphics[width=\linewidth]{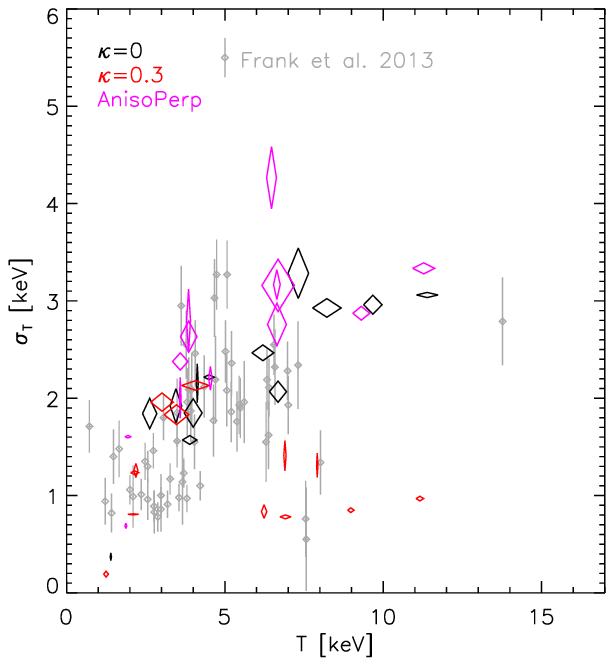}
		\caption{Comparing the temperature fluctuations within $R_{2500}$ as inferred from observations by \protect\cite{Frank2013} with the one predicted for the set of simulated clusters, including also some less massive ones which are found within the high resolution region of the zoomed simulations.
		The different colors correspond to the simulations without thermal conduction (black), isotropic thermal conduction with $\kappa = 0.3$ (red) and anisotropic thermal conduction, where the perpendicular term is evaluated using the linear scaling (pink).}
		\label{FIGcomp2}
	\end{figure}
	
	{At last, again, we compare the temperature fluctuations to the observational data of \cite{Frank2013}.
	Similar as before, beside the central, massive cluster we also take other clusters found in the high resolution region	into account, allowing us to get also some objects with various temperatures, sampling the low temperature region.
	Fig. \ref{FIGcomp2} shows the comparison of the data with our simulations without thermal	conduction, with isotropic conduction using $\kappa = 0.3$ and with	anisotropic conduction including linear scaling for the perpendicular component.
	While the simulated clusters with isotropic conduction using $\kappa=0.3$ fall significantly below the bulk of data points for clusters above 5 keV, the simulations without thermal conduction and with anisotropic thermal conduction seem to represent the observed data points reasonably well.
	Still a more clearly selected set of simulated galaxy clusters across the whole temperature range as well as observations	in the high temperature regime are needed to draw more robust conclusions.

\subsection{The magnetic field structure in the simulated clusters}

	\begin{figure}
		\centering
		\includegraphics[width=\linewidth]{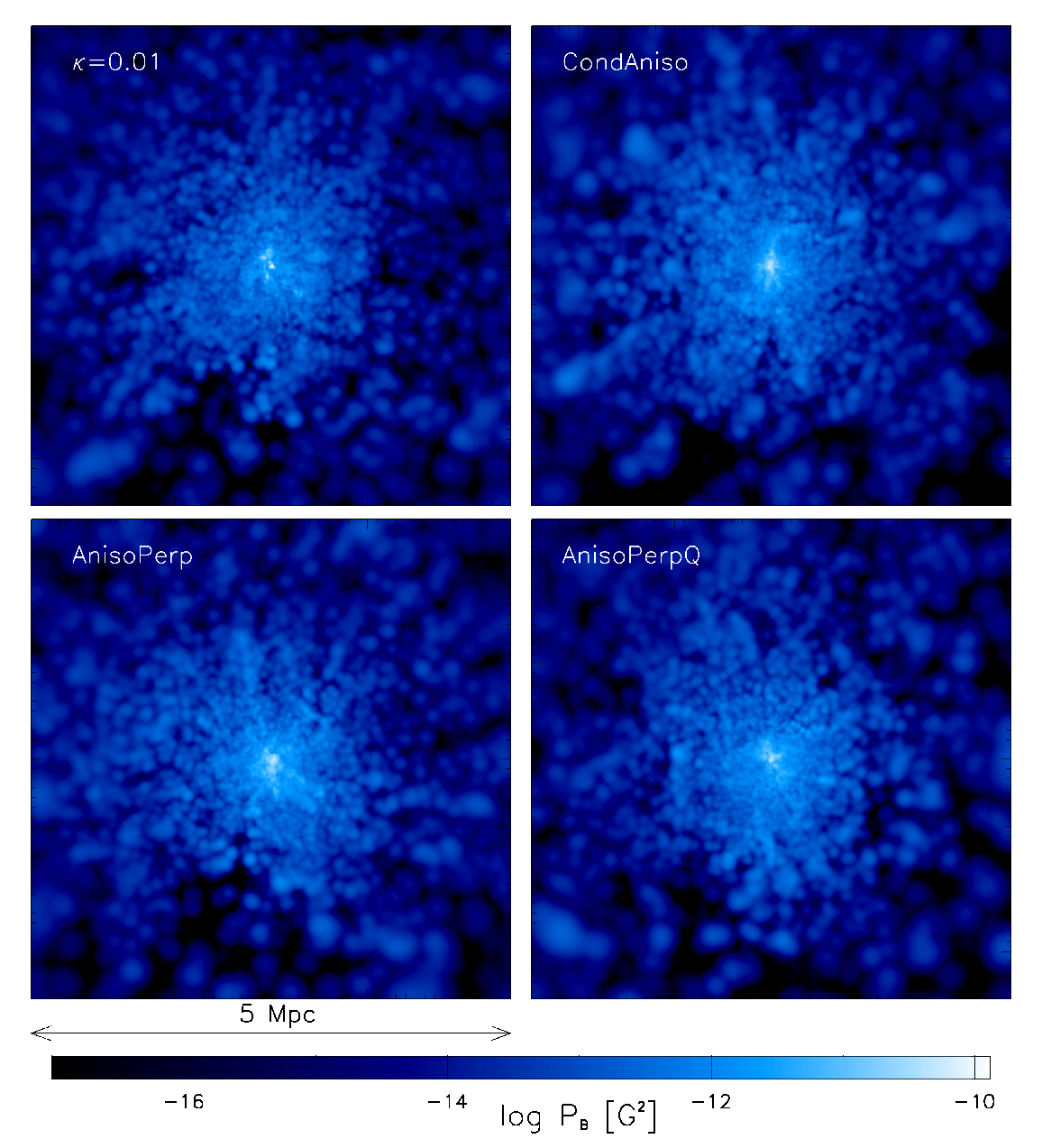}
		\caption{We present maps of the magnetic pressure $P_B = B^2 / 8 \pi$ in a thin slice through the cluster centre. As before we compare four runs with different conduction settings.}
		\label{FIGmagmaps}
	\end{figure}
	
	Since there is a tight connection between thermal conduction and magnetic field evolution, we now investigate the magnetic field in the simulated galaxy clusters. We show in Fig. \ref{FIGmagmaps} thin slices of the magnetic pressure through the cluster centre.
	In high density regions, in the cores of clusters and groups, almost all particles have magnetic fields at $\mu$G levels, while they drop down to nG levels in the outer regions. Since the magnetic fields are introduced by supernova seeding events we see, that the high magnetic fields are strongly localised in our simulations. Transport processes smooth out the magnetic field distribution but a bubbly structure still remains. Particles which are not directly influenced by the supernova seeding have rather low magnetic field values as low as even $B \approx 10^{-20} \; \mathrm{G}$. These particles are sensitive to the chosen scaling of the conduction mechanism.
		
	Still, also particles sitting in extreme density peaks contribute significantly to this dependency, since Eqs. \ref{EQdiffusionbsquared} and \ref{EQdiffusionblinear} are not only dependent on the magnetic field strength but also on density. Therefore this discrimination of scaling is always important and not only an artefact of the magnetic field seeding mechanism.
	
	\begin{figure}
		\centering
		\includegraphics[angle=270,width=\linewidth]{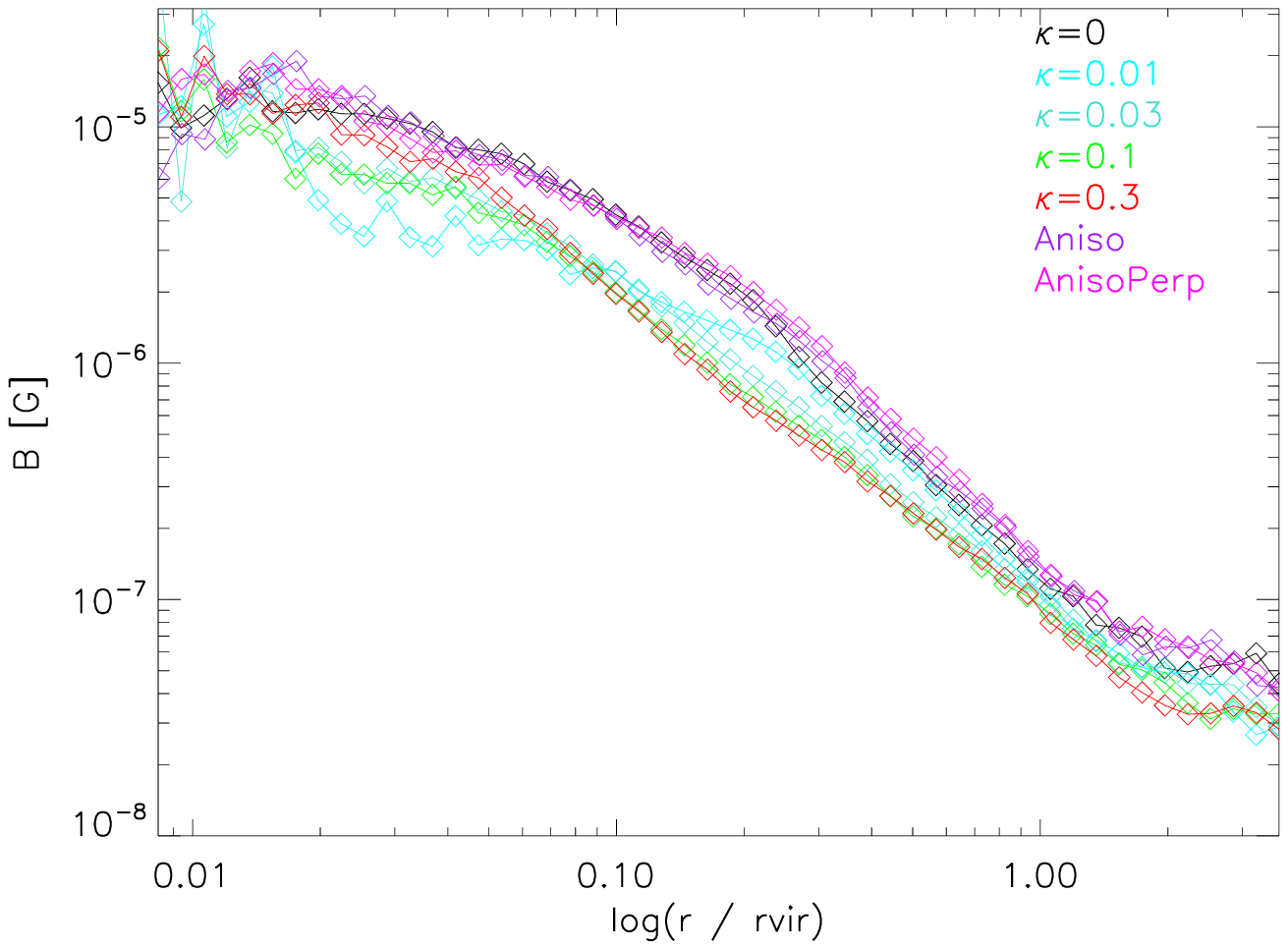}
		\caption{We show the volume-averaged radial magnetic field profiles of the most massive galaxy cluster for different conduction settings. Stronger conduction leads to a steeper profile while the curves of the anisotropic runs are closest to the non-conductive run.}
		\label{FIGmagrad}
	\end{figure}
	
	Since differences in the four maps are only marginally, we also present the radial magnetic field profiles (volume-averaged) for several runs in Fig. \ref{FIGmagrad}. We see that the radial profiles for all different runs are of very similar shape and broadly agree with previous findings in the literature. The more efficient we allow conduction to be, the steeper the profiles become outside the cluster core. The runs with anisotropic conduction resemble again runs with a very low isotropic coefficient.

	\begin{figure}
		\centering
		\includegraphics[angle=270,width=\linewidth]{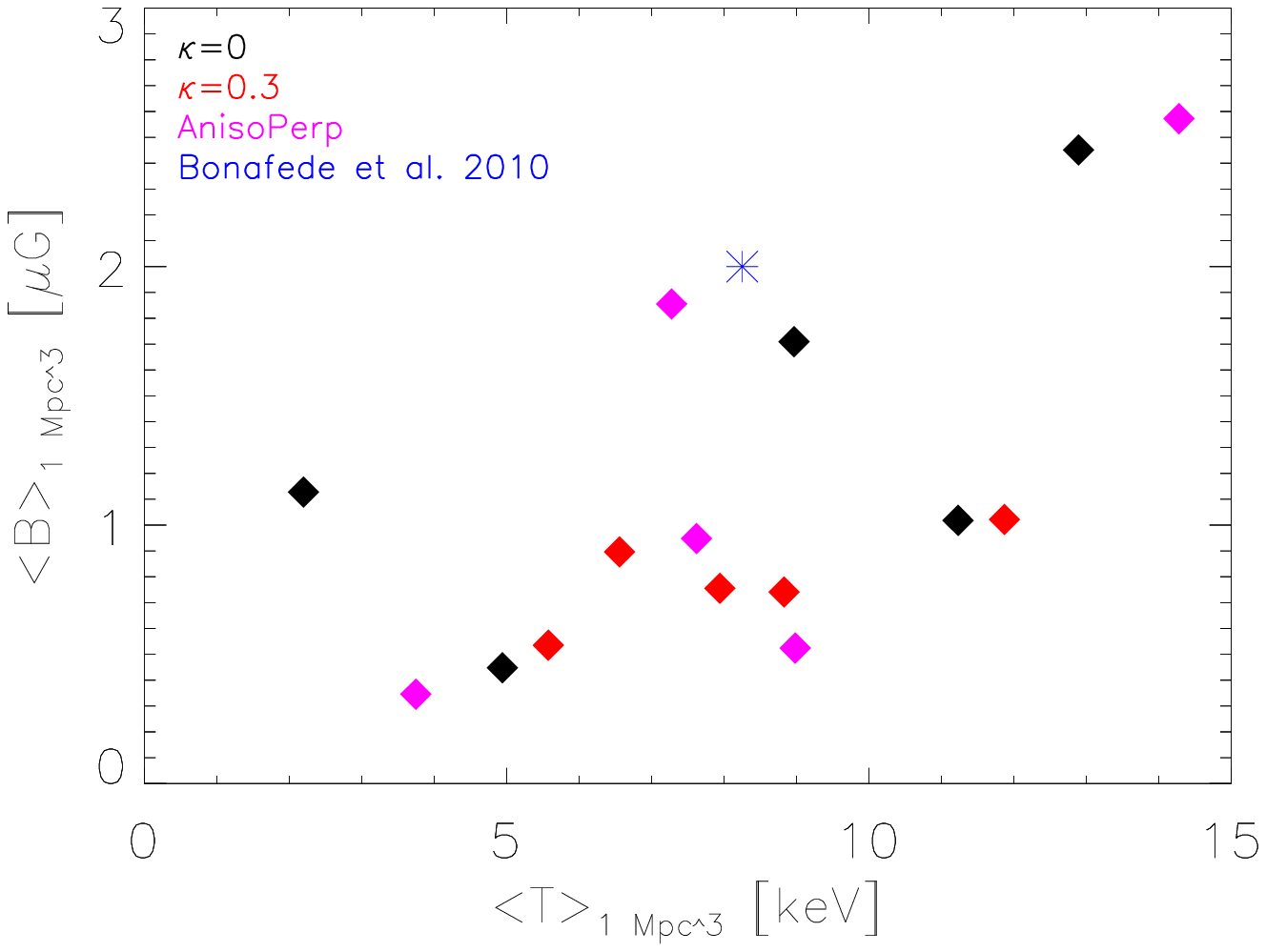}
		\caption{We show the mean volume-averaged temperature against the mean volume-averaged magnetic field strength of the five most massive clusters inside the central one Mpc$^3$. Additionally, we overplot an observational data point for the Coma Cluster taken from \citet{Bonafede2010a}.}
		\label{FIGbovert}
	\end{figure}
	
	To illustrate the resulting magnetic field strengths in the larger cluster sample we plot temperature against magnetic field volume-averaged within a sphere of volume one Mpc$^3$ around the cluster centres in Fig. \ref{FIGbovert}. We choose this region in order to compare to an observational result for the Coma cluster presented by \citet{Bonafede2010a}. Overall, the results obtained with our simulated clusters are in good agreement with the magnetic field observed in the Coma cluster, although there is a trend for some clusters to have slightly lower magnetic fields. The amplitudes are more in the range of mean magnetic fields inferred from the radio halo \citep[e.g. $0.7-1.9 \mu\mathrm{G}$ given by][]{Thierbach2003} than from rotation measurements. Compared to previously findings, this is probably related to our seeding mechanism by supernova events.

	Comparing the different runs we present, one can clearly see that changes in the conduction prescription pose an overall influence over the presented averaged quantities. The runs without thermal conduction and with our anisotropic prescription produce a larger variance of cluster temperatures and higher magnetic fields on the high temperature end in comparison to runs with rather strong isotropic thermal conduction. This matches well with the conclusions we drew from the radial temperature profiles in the previous subsection.

\section{Summary and Conclusions}

	We derive and discuss a numerical scheme for anisotropic thermal conduction in the presence of magnetic fields.
	We present a discretisation for SPH and implement our new method into the cosmological simulation code GADGET.
	We show a variety of standard tests as well as cosmological simulations of galaxy cluster formation with different choice of conduction parameters, where we combined the new conduction implementation with a supernova seeding scheme for the magnetic field (Beck et al. 2013), leading to a self consistent evolution of magnetic fields within the cosmic structures.
	
	Our numerical scheme for anisotropic conduction in SPH solves the corresponding equations using a conjugate gradient solver, and therefore need only a very small amount of extra computational effort in the MHD version of GADGET.
	However, the straight forward derivation can violate the second law of thermodynamics in cases of strong anisotropies and large jumps in temperature.
	Additionally, this can causes an unstable behaviour of the conjugate gradient solver in the presence of extremely sharp jumps of temperature.
	Typically, this problem is solved by introducing a correction which ensures positive definiteness of the linear equation system by adding an artificial isotropic component.
	However, this correction can lead to significant, artificial heat flow perpendicular to the magnetic field and it is therefore questionable if such a numerical correction is useful in a realistic environment, where it can hide the effect of anisotropic conduction.
	In general, for any realistic situation, our anisotropic implementation with fully suppressed perpendicular term is already stable enough so that we do not have to add an artificial, destructive isotropisation term.
	However, a closer look at the term of the perpendicular conduction coefficient reveals that the amount of suppression of the perpendicular conduction coefficients can scale with either with $B^{-1}$ or with $B^{-2}$ \citep{Huba2011}.
	Depending on other plasma properties, these two scalings can have different relative effect on the amount of perpendicular heat transport.
	To test them, we perform cosmological simulations of the formation of galaxy clusters with different implementations of the perpendicular transport coefficients.
	We also compare the results with a fully isotropic implementation of heat transport for different values of the suppression of thermal conduction with respect to the classical Spitzer value.
	Our main results can be summarised as follows:
	
	\begin{itemize}
		\item Temperature maps from simulated galaxy clusters show that isotropic thermal conduction not only transports heat outwards, but also smoothes small-scale features.
		Anisotropic conduction seems to resemble isotropic transport with coefficients like $\kappa \sim 0.01 \cdot \kappa_\mathrm{Sp}$; however, prominent substructures in the temperature distribution survive due to insulation by magnetic field lines.
		\item Radial temperature profiles change differently when applying anisotropic thermal conduction depending on the dynamical evolution of the cluster. Some profiles are very similar to those without thermal conduction showing a rising profile with a large drop towards the center while others show a very pronounced cool core. In contrast, isotropic conduction produce flattened, almost isothermal temperature profiles in the central regions.
		\item We show the relevance of a proper treatment of perpendicular conduction instead of only parallel transport.
		At all times we find a significant amount of particles with temperatures $T > 10^7$ K for which a non negligible perpendicular component is assigned.
		These particles sit either in regions with negligible magnetic field or at gas density peaks.
		The particles with low or no magnetic field are especially important in the simulations, since many particles have not undergone magnetic supernova seeding events.
		\item We find that different conduction prescriptions also have some influence on the resulting magnetic field of the clusters.
		Lower conduction leads to steeper radial profiles in the most massive cluster.
		In the larger sample we see that anisotropic conduction leads to a larger variance of mean cluster temperatures and slightly stronger magnetic fields on the high temperature end.
		\item We calculate emissivity distributions and compare to observed temperature fluctuations of \cite{Frank2013}.
		We find that simulations with either zero or anisotropic conduction reflect the observational data points best, whereas isotropic conduction with $\kappa > 0.1$ shows a clear lack of temperature fluctuations compared to the observational data points.
		In comparison, although clearly visible, the differences for the different descriptions for perpendicular conduction show only mild changes in the amount of temperature fluctuations. Here a significant increase of number of simulated galaxy clusters as well as many more observations of high temperature clusters will be needed to discriminate between them.
		\item We compare the fractions of cold gas and stars in different simulations and find a weak dependence on the conduction parameters.
		Conduction seems not to play a key role in suppressing cooling in galaxy clusters, but due to the coupling of the suppression factors to the local dynamical state of the cluster, the anisotropic conduction might contribute to the observed bimodality of cool core and non cool core systems.
	\end{itemize}
	
	In conclusion, anisotropic thermal conduction is not a dynamically dominant process within galaxy clusters, but can influence the evolution of small-scale structure.
	In contrary to isotropic heat conduction, it produces a reasonable amount of temperature fluctuations compared to observations and still allows locally for transport of heat.
	In general, it comes only with a small amount of computational cost for cosmological SPMHD codes and eliminates the need for a free efficiency parameter.
	A next step would be to consider also non-collisional thermal conduction and include the cross term mentioned in section \ref{anisotropicthermalconduction}. Since we have seen that a perpendicular conduction component can have quite an impact we expect this term to also play a role in the process.
	
	In the future, a larger sample of cosmological simulations with possibly even higher resolution will allow a better statistical analysis of the detailed temperature and magnetic field structures and the role of anisotropic transport effects in galaxy clusters.
	This will help to gain further knowledge about the scaling relations of perpendicular conduction or the importance of small-scale plasma instabilities, which is very challenging to resolve.

\section*{Acknowledgements}
We thank Volker Springel for access to the developer version of GADGET.
\nr{We thank the anonymous referees for very helpful comments that helped to further improve this work along the way.}
AA gives special thanks to his colleagues Max Imgrund, Marco H\"auser and Carsten Uphoff for very useful discussions during the creation of this work.
KD and AMB are supported by the DFG Research Unit 1254 'Magnetisation of interstellar and intergalactic media' and by the DFG Cluster of Excellence ‘Origin and Structure of the Universe’.
KD is supported by the SFB-Transregio TR33 'The Dark Universe'.
We also thank the anonymous referee for very helpful comments to this article.
AA, KD, AMB and MP are part of the MAGNETICUM\footnote{http://www.magneticum.org} simulation project.

\bibliography{sources}

\begin{thebibliography}{}
\makeatletter
\relax
\def\mn@urlcharsother{\let\do\@makeother \do\$\do\&\do\#\do\^\do\_\do\%\do\~}
\def\mn@doi{\begingroup\mn@urlcharsother \@ifnextchar [ {\mn@doi@}
  {\mn@doi@[]}}
\def\mn@doi@[#1]#2{\def\@tempa{#1}\ifx\@tempa\@empty \href
  {http://dx.doi.org/#2} {doi:#2}\else \href {http://dx.doi.org/#2} {#1}\fi
  \endgroup}
\def\mn@eprint#1#2{\mn@eprint@#1:#2::\@nil}
\def\mn@eprint@arXiv#1{\href {http://arxiv.org/abs/#1} {{\tt arXiv:#1}}}
\def\mn@eprint@dblp#1{\href {http://dblp.uni-trier.de/rec/bibtex/#1.xml}
  {dblp:#1}}
\def\mn@eprint@#1:#2:#3:#4\@nil{\def\@tempa {#1}\def\@tempb {#2}\def\@tempc
  {#3}\ifx \@tempc \@empty \let \@tempc \@tempb \let \@tempb \@tempa \fi \ifx
  \@tempb \@empty \def\@tempb {arXiv}\fi \@ifundefined
  {mn@eprint@\@tempb}{\@tempb:\@tempc}{\expandafter \expandafter \csname
  mn@eprint@\@tempb\endcsname \expandafter{\@tempc}}}

\bibitem[\protect\citeauthoryear{Andreon}{Andreon}{2010}]{Andreon2010}
Andreon S.,  2010, \mn@doi [MNRAS] {10.1111/j.1365-2966.2010.16856.x}, 407, 263

\bibitem[\protect\citeauthoryear{Avara, Reynolds  \& Bogdanovi\'{c}}{Avara
  et~al.}{2013}]{Avara2013}
Avara M.~J.,  Reynolds C.~S.,   Bogdanovi\'{c} T.,  2013, preprint (\mn@eprint
  {arXiv} {arXiv:1305.0281v1})

\bibitem[\protect\citeauthoryear{Balogh, Pearce, Bower  \& Kay}{Balogh
  et~al.}{2001}]{Balogh2001}
Balogh M.~L.,  Pearce F.~R.,  Bower R.~G.,   Kay S.~T.,  2001, \mn@doi [MNRAS]
  {10.1111/j.1365-2966.2001.04667.x}, 326, 1228

\bibitem[\protect\citeauthoryear{Beck, Lesch, Dolag, Kotarba, Geng  \&
  Stasyszyn}{Beck et~al.}{2012}]{Beck2012}
Beck A.~M.,  Lesch H.,  Dolag K.,  Kotarba H.,  Geng A.,   Stasyszyn F.,  2012,
  \mn@doi [MNRAS] {10.1111/j.1365-2966.2012.20759.x}, 422, 2152

\bibitem[\protect\citeauthoryear{Beck, Dolag, Lesch  \& Kronberg}{Beck
  et~al.}{2013}]{Beck2013}
Beck A.~M.,  Dolag K.,  Lesch H.,   Kronberg P.~P.,  2013, \mn@doi [MNRAS]
  {10.1093/mnras/stt1549}, 435, 3575

\bibitem[\protect\citeauthoryear{Beck et~al.,}{Beck et~al.}{2016}]{Beck2015}
Beck A.~M.,  et~al., 2016, \mn@doi [MNRAS] {10.1093/mnras/stv2443}, 455, 2110

\bibitem[\protect\citeauthoryear{Binney \& Cowie}{Binney \&
  Cowie}{1981}]{Binney1981}
Binney J.,  Cowie L.~L.,  1981, ApJ, 247, 464

\bibitem[\protect\citeauthoryear{Bogdanovi\'{c}, Reynolds, Balbus  \&
  Parrish}{Bogdanovi\'{c} et~al.}{2009}]{Bogdanovic2009}
Bogdanovi\'{c} T.,  Reynolds C.~S.,  Balbus S.~A.,   Parrish I.~J.,  2009,
  \mn@doi [ApJ] {10.1088/0004-637X/704/1/211}, 704, 211

\bibitem[\protect\citeauthoryear{Bonafede, Feretti, Murgia, Govoni, Giovannini,
  Dallacasa, Dolag  \& Taylor}{Bonafede et~al.}{2010}]{Bonafede2010a}
Bonafede A.,  Feretti L.,  Murgia M.,  Govoni F.,  Giovannini G.,  Dallacasa
  D.,  Dolag K.,   Taylor G.~B.,  2010, \mn@doi [A\&A]
  {10.1051/0004-6361/200913696}, 513, A30

\bibitem[\protect\citeauthoryear{Bonafede, Dolag, Stasyszyn, Murante  \&
  Borgani}{Bonafede et~al.}{2011}]{Bonafede2011}
Bonafede A.,  Dolag K.,  Stasyszyn F.,  Murante G.,   Borgani S.,  2011,
  \mn@doi [MNRAS] {10.1111/j.1365-2966.2011.19523.x}, 418, 2234

\bibitem[\protect\citeauthoryear{Braginskii}{Braginskii}{1965}]{Braginskii1965}
Braginskii S.~I.,  1965, in , Review of Plasma Physics.
p.~205

\bibitem[\protect\citeauthoryear{Bregman \& David}{Bregman \&
  David}{1989}]{Bregman1989}
Bregman J.~N.,  David L.~P.,  1989, ApJ, 341, 49

\bibitem[\protect\citeauthoryear{Brookshaw}{Brookshaw}{1985}]{Brookshaw1985}
Brookshaw L.,  1985, in Astronomical Society of Australia, Proceeding. pp
  207--210

\bibitem[\protect\citeauthoryear{Chandran \& Cowley}{Chandran \&
  Cowley}{1998}]{Chandran1998}
Chandran B.,  Cowley S.,  1998, \mn@doi [Phys. Rev. Lett.]
  {10.1103/PhysRevLett.80.3077}, 80, 3077

\bibitem[\protect\citeauthoryear{Cleary \& Monaghan}{Cleary \&
  Monaghan}{1999}]{Cleary1999}
Cleary P.~W.,  Monaghan J.~J.,  1999, \mn@doi [J. Comp. Phys.]
  {10.1006/jcph.1998.6118}, 148, 227

\bibitem[\protect\citeauthoryear{Cowie \& McKee}{Cowie \&
  McKee}{1977}]{CowieLennoxL.McKee1977}
Cowie L.~L.,  McKee C.~F.,  1977, ApJ, 211, 135

\bibitem[\protect\citeauthoryear{Dehnen \& Aly}{Dehnen \&
  Aly}{2012}]{Dehnen2012}
Dehnen W.,  Aly H.,  2012, \mn@doi [MNRAS] {10.1111/j.1365-2966.2012.21439.x},
  1082

\bibitem[\protect\citeauthoryear{Dolag \& Stasyszyn}{Dolag \&
  Stasyszyn}{2009}]{Dolag2009}
Dolag K.,  Stasyszyn F.,  2009, \mn@doi [MNRAS]
  {10.1111/j.1365-2966.2009.15181.x}, 398, 1678

\bibitem[\protect\citeauthoryear{Dolag, Bartelmann  \& Lesch}{Dolag
  et~al.}{1999}]{Dolag1999}
Dolag K.,  Bartelmann M.,   Lesch H.,  1999, A\&A, 363, 4

\bibitem[\protect\citeauthoryear{Dolag, Jubelgas, Springel, Borgani  \&
  Rasia}{Dolag et~al.}{2004}]{Dolag2004}
Dolag K.,  Jubelgas M.,  Springel V.,  Borgani S.,   Rasia E.,  2004, ApJ, 606,
  2000

\bibitem[\protect\citeauthoryear{Dubois \& Commer\c{c}on}{Dubois \&
  Commer\c{c}on}{2015}]{Dubois2015}
Dubois Y.,  Commer\c{c}on B.,  2015, preprint (\mn@eprint {arXiv}
  {arXiv:1509.07037v3})

\bibitem[\protect\citeauthoryear{Fabian}{Fabian}{1994}]{Fabian1994}
Fabian A.~C.,  1994, \mn@doi [Ann. Rev. of A\&A]
  {10.1146/annurev.astro.32.1.277}, 32, 277

\bibitem[\protect\citeauthoryear{Fabian}{Fabian}{2002}]{Fabian2002}
Fabian A.~C.,  2002, in Lighthouses of the Universe: The Most Luminous
  Celestial Objects and Their Use for Cosmology.  (\mn@eprint {arXiv}
  {0201386v1}), \mn@doi{10.1007/10856495\_3}

\bibitem[\protect\citeauthoryear{Fabian, Sanders, Taylor, Allen, Crawford,
  Johnstone  \& Iwasawa}{Fabian et~al.}{2006}]{Fabian2006}
Fabian A.~C.,  Sanders J.~S.,  Taylor G.~B.,  Allen S.~W.,  Crawford C.~S.,
  Johnstone R.~M.,   Iwasawa K.,  2006, \mn@doi [MNRAS]
  {10.1111/j.1365-2966.2005.09896.x}, 366, 417

\bibitem[\protect\citeauthoryear{Frank-Kamenezki}{Frank-Kamenezki}{1967}]{Frank-Kamenezki1967}
Frank-Kamenezki D.~A.,  1967, {Vorlesungen \"{u}ber Plasmaphysik}.
VEB Deutscher Verlag der Wissenschaften

\bibitem[\protect\citeauthoryear{Frank, Peterson, Andersson, Fabian  \&
  Sanders}{Frank et~al.}{2013}]{Frank2013}
Frank K.~a.,  Peterson J.~R.,  Andersson K.,  Fabian a.~C.,   Sanders J.~S.,
  2013, \mn@doi [ApJ] {10.1088/0004-637X/764/1/46}, 764, 46

\bibitem[\protect\citeauthoryear{Gaspari, Churazov, Nagai, Lau  \&
  Zhuravleva}{Gaspari et~al.}{2014}]{Gaspari2014}
Gaspari M.,  Churazov E.,  Nagai D.,  Lau E.~T.,   Zhuravleva I.,  2014, A\&A,
  67, 1

\bibitem[\protect\citeauthoryear{Golant, Zhilinsky  \& Sakharov}{Golant
  et~al.}{1980}]{Golant1980}
Golant V.~E.,  Zhilinsky A.~P.,   Sakharov I.~E.,  1980, {Fundamentals of
  Plasma Physics}.
John Wiley \& Sons Inc

\bibitem[\protect\citeauthoryear{Guthrie \& Wakerling}{Guthrie \&
  Wakerling}{1949}]{Guthrie1949}
Guthrie A.,  Wakerling R.~K.,  1949, {The Characteristics of Electrical
  Discharges in Magnetic Fields}.
National nuclear energy series: Electromagnetic Separation Project, McGraw-Hill

\bibitem[\protect\citeauthoryear{Hopkins}{Hopkins}{2016}]{Hopkins2016}
Hopkins P.~F.,  2016, preprint, 000 (\mn@eprint {arXiv} {1602.07703})

\bibitem[\protect\citeauthoryear{Huba}{Huba}{2011}]{Huba2011}
Huba J.~D.,  2011, {NRL PLASMA FORMULARY Supported by The Office of Naval
  Research}

\bibitem[\protect\citeauthoryear{Jubelgas, Springel  \& Dolag}{Jubelgas
  et~al.}{2004}]{Jubelgas2004}
Jubelgas M.,  Springel V.,   Dolag K.,  2004, \mn@doi [MNRAS]
  {10.1111/j.1365-2966.2004.07801.x}, 351, 423

\bibitem[\protect\citeauthoryear{Kannan, Springel, Pakmor, Marinacci  \&
  Vogelsberger}{Kannan et~al.}{2015}]{Kannan2015}
Kannan R.,  Springel V.,  Pakmor R.,  Marinacci F.,   Vogelsberger M.,  2015,
  preprint, 17, 1 (\mn@eprint {arXiv} {1512.03053})

\bibitem[\protect\citeauthoryear{Koerner, Portsmouth, Sadlo, Ertl  \&
  Eberhardt}{Koerner et~al.}{2014}]{Koerner2014}
Koerner D.,  Portsmouth J.,  Sadlo F.,  Ertl T.,   Eberhardt B.,  2014, \mn@doi
  [Computer Graphics Forum] {10.1111/cgf.12342}, 33, 178

\bibitem[\protect\citeauthoryear{Komarov, Churazov  \& Schekochihin}{Komarov
  et~al.}{2014a}]{Komarov2013}
Komarov S.,  Churazov E.,   Schekochihin A.~A.,  2014a, MNRAS, 440, 1153

\bibitem[\protect\citeauthoryear{Komarov, Churazov, Schekochihin  \&
  ZuHone}{Komarov et~al.}{2014b}]{Komarov2014}
Komarov S.~V.,  Churazov E.~M.,  Schekochihin a.~a.,   ZuHone J.~a.,  2014b,
  \mn@doi [MNRAS] {10.1093/mnras/stu281}, 440, 1153

\bibitem[\protect\citeauthoryear{Kravtsov, Vikhlinin  \&
  Meshscheryakov}{Kravtsov et~al.}{2014}]{Kravtsov2014}
Kravtsov A.,  Vikhlinin A.,   Meshscheryakov A.,  2014, preprint, p.~21
  (\mn@eprint {arXiv} {1401.7329})

\bibitem[\protect\citeauthoryear{Kronberg}{Kronberg}{1994}]{Kronberg1994}
Kronberg P.~P.,  1994, Reports on Progress in Physics, 57, 325

\bibitem[\protect\citeauthoryear{Landau \& Lifschitz}{Landau \&
  Lifschitz}{2007}]{Landau2007}
Landau L.~D.,  Lifschitz E.~M.,  2007, {Lehrbuch der theoretischen Physik VI
  Hydrodynamik}, 5th edn.
Harri Deutsch GmbH, Moskau

\bibitem[\protect\citeauthoryear{Lin, Mohr  \& Stanford}{Lin
  et~al.}{2003}]{Lin2003}
Lin Y.-T.,  Mohr J.~J.,   Stanford S.~A.,  2003, ApJ, 591, 749

\bibitem[\protect\citeauthoryear{Loeb}{Loeb}{2002}]{Loeb2002}
Loeb A.,  2002, New Astron., 7, 279

\bibitem[\protect\citeauthoryear{Narayan \& Medvedev}{Narayan \&
  Medvedev}{2001}]{Narayan2001}
Narayan R.,  Medvedev M.~V.,  2001, \mn@doi [ApJ] {10.1086/338325}, 562, 129

\bibitem[\protect\citeauthoryear{Owers, Nulsen, Couch  \& Markevitch}{Owers
  et~al.}{2009}]{Owers2009}
Owers M.~S.,  Nulsen P. E.~J.,  Couch W.~J.,   Markevitch M.,  2009, \mn@doi
  [ApJ] {10.1088/0004-637X/704/2/1349}, 704, 1349

\bibitem[\protect\citeauthoryear{Owers, Nulsen  \& Couch}{Owers
  et~al.}{2011}]{Owers2011}
Owers M.~S.,  Nulsen P. E.~J.,   Couch W.~J.,  2011, \mn@doi [ApJ]
  {10.1088/0004-637X/741/2/122}, 741, 122

\bibitem[\protect\citeauthoryear{Parrish \& Stone}{Parrish \&
  Stone}{2005}]{Parrish2005}
Parrish I.~J.,  Stone J.~M.,  2005, \mn@doi [ApJ] {10.1086/444589}, 633, 334

\bibitem[\protect\citeauthoryear{Parrish, Quataert  \& Sharma}{Parrish
  et~al.}{2009}]{Parrish2009}
Parrish I.~J.,  Quataert E.,   Sharma P.,  2009, \mn@doi [ApJ]
  {10.1088/0004-637X/703/1/96}, p.~14

\bibitem[\protect\citeauthoryear{Peterson \& Fabian}{Peterson \&
  Fabian}{2006}]{Peterson2006}
Peterson J.~R.,  Fabian A.~C.,  2006, \mn@doi [Phys. Rep.]
  {10.1016/j.physrep.2005.12.007}, 427, 1

\bibitem[\protect\citeauthoryear{Peterson, Kahn, Paerels, Kaastra, Tamura,
  Bleeker, Ferrigno  \& Jennigan}{Peterson et~al.}{2003}]{Peterson2003}
Peterson J.~R.,  Kahn S.~M.,  Paerels F. B.~S.,  Kaastra J.~S.,  Tamura T.,
  Bleeker J. A.~M.,  Ferrigno C.,   Jennigan J.,  2003, ApJ, 590, 207

\bibitem[\protect\citeauthoryear{Petkova \& Springel}{Petkova \&
  Springel}{2009}]{Petkova2009}
Petkova M.,  Springel V.,  2009, \mn@doi [MNRAS]
  {10.1111/j.1365-2966.2009.14843.x}, 396, 1383

\bibitem[\protect\citeauthoryear{Pistinner \& Shaviv}{Pistinner \&
  Shaviv}{1996}]{Pistinner1996}
Pistinner S.,  Shaviv G.,  1996, ApJ, 459, 147

\bibitem[\protect\citeauthoryear{Price}{Price}{2012}]{Price2012}
Price D.~J.,  2012, \mn@doi [J. Comp. Phys.] {10.1016/j.jcp.2010.12.011}, 231,
  759

\bibitem[\protect\citeauthoryear{Rasera \& Chandran}{Rasera \&
  Chandran}{2008}]{Rasera2008}
Rasera Y.,  Chandran B.,  2008, \mn@doi [ApJ] {10.1086/591012}, p.~40

\bibitem[\protect\citeauthoryear{Rasia et~al.,}{Rasia et~al.}{2014}]{Rasia2014}
Rasia E.,  et~al., 2014, \mn@doi [ApJ] {10.1088/0004-637X/791/2/96}, 791, 96

\bibitem[\protect\citeauthoryear{Rasia et~al.,}{Rasia et~al.}{2015}]{Rasia2015}
Rasia E.,  et~al., 2015, ApJ Letters, 813

\bibitem[\protect\citeauthoryear{Rechester \& Rosenbluth}{Rechester \&
  Rosenbluth}{1978}]{Rechester1978}
Rechester A.~B.,  Rosenbluth M.,  1978, Phys. Rev. Lett., 40, 38

\bibitem[\protect\citeauthoryear{Rosner \& Tucker}{Rosner \&
  Tucker}{1989}]{Rosner1989}
Rosner R.,  Tucker W.~H.,  1989, \mn@doi [ApJ] {10.1086/167234}, 338, 761

\bibitem[\protect\citeauthoryear{Ruderman, Oliver, Erd\'{e}lyi, Ballester  \&
  Goossens}{Ruderman et~al.}{2000}]{Ruderman2000}
Ruderman M.,  Oliver R.,  Erd\'{e}lyi R.,  Ballester J.,   Goossens M.,  2000,
  Astronomy \& Astrophysics, 276, 261

\bibitem[\protect\citeauthoryear{Ruszkowski, Lee, Br\"{u}ggen, Parrish  \&
  Oh}{Ruszkowski et~al.}{2011}]{Ruszkowski2011}
Ruszkowski M.,  Lee D.,  Br\"{u}ggen M.,  Parrish I.~J.,   Oh S.~P.,  2011,
  \mn@doi [ApJ] {10.1088/0004-637X/740/2/81}, 740, 81

\bibitem[\protect\citeauthoryear{Saad}{Saad}{2003}]{Saad2000}
Saad Y.,  2003, {Iterative Methods for Sparse Linear Systems}, 2nd edn.
Society for Industrial and Applied Mathematics, \url
  {http://www-users.cs.umn.edu/~saad/books.html}

\bibitem[\protect\citeauthoryear{Saad \& Schultz}{Saad \&
  Schultz}{1986}]{Saad1986}
Saad Y.,  Schultz M.~H.,  1986, SIAM J. ScI. STAT. COMPUT. Vo|, 7, 856

\bibitem[\protect\citeauthoryear{Sarazin}{Sarazin}{1986}]{Sarazin1986}
Sarazin C.,  1986, \mn@doi [Rev. Mod. Phys.] {10.1103/RevModPhys.58.1}, 58, 1

\bibitem[\protect\citeauthoryear{Sarazin}{Sarazin}{2008}]{Sarazin2008}
Sarazin C.,  2008, {Gas Dynamics in Clusters of Galaxies},
  \mn@doi{10.1007/978-1-4020-6941-3}, \url
  {http://www.springerlink.com/index/10.1007/978-1-4020-6941-3}

\bibitem[\protect\citeauthoryear{Schekochihin, Cowley, Kulsrud, Rosin  \&
  Heinemann}{Schekochihin et~al.}{2008}]{Schekochihin2008}
Schekochihin A.~A.,  Cowley S.~C.,  Kulsrud R.~M.,  Rosin M.~S.,   Heinemann
  T.,  2008, \mn@doi [Phys. Rev. Lett.] {10.1103/PhysRevLett.100.081301}, 100,
  1

\bibitem[\protect\citeauthoryear{Sharma \& Hammett}{Sharma \&
  Hammett}{2007}]{Sharma2007}
Sharma P.,  Hammett G.~W.,  2007, \mn@doi [J. Comp. Phys.]
  {10.1016/j.jcp.2007.07.026}, 227, 123

\bibitem[\protect\citeauthoryear{Sharma, Parrish  \& Quataert}{Sharma
  et~al.}{2010}]{Sharma2010}
Sharma P.,  Parrish I.~J.,   Quataert E.,  2010, \mn@doi [ApJ]
  {10.1088/0004-637X/720/1/652}, p.~31

\bibitem[\protect\citeauthoryear{Sonneveld}{Sonneveld}{1989}]{Sonneveld1989}
Sonneveld P.,  1989, SIAM J. Sci. Statist. Comput., 10, 35

\bibitem[\protect\citeauthoryear{Spitzer}{Spitzer}{1956}]{Spitzer1956}
Spitzer L.~J.,  1956, {Physics of fully ionized gas}.
Interscience Publishers, \url {http://books.google.de/books?id=CilRAAAAMAAJ}

\bibitem[\protect\citeauthoryear{Spitzer \& H\"{a}rm}{Spitzer \&
  H\"{a}rm}{1953}]{Spitzer1953}
Spitzer L.~J.,  H\"{a}rm R.,  1953, \mn@doi [Phys. Rev. Lett.]
  {10.1103/PhysRev.89.977}, 89, 977

\bibitem[\protect\citeauthoryear{Springel}{Springel}{2005}]{Springel2005a}
Springel V.,  2005, \mn@doi [MNRAS] {10.1111/j.1365-2966.2005.09655.x}, 364,
  1105

\bibitem[\protect\citeauthoryear{Springel \& Hernquist}{Springel \&
  Hernquist}{2002}]{Springel2002}
Springel V.,  Hernquist L.,  2002, MNRAS, 333, 649

\bibitem[\protect\citeauthoryear{Springel \& Hernquist}{Springel \&
  Hernquist}{2003}]{Springel2003}
Springel V.,  Hernquist L.,  2003, \mn@doi [MNRAS]
  {10.1046/j.1365-8711.2003.06206.x}, 339, 289

\bibitem[\protect\citeauthoryear{Springel, Yoshida  \& White}{Springel
  et~al.}{2001}]{Springel2001}
Springel V.,  Yoshida N.,   White S.~D.,  2001, \mn@doi [New Astron.]
  {10.1016/S1384-1076(01)00042-2}, 6, 79

\bibitem[\protect\citeauthoryear{Stasyszyn, Dolag  \& Beck}{Stasyszyn
  et~al.}{2013}]{Stasyszyn2013}
Stasyszyn F.,  Dolag K.,   Beck A.~M.,  2013, MNRAS, 428

\bibitem[\protect\citeauthoryear{Suzuki, Ogawa, Matsumoto  \& Matsumoto}{Suzuki
  et~al.}{2013}]{Suzuki2013}
Suzuki K.,  Ogawa T.,  Matsumoto Y.,   Matsumoto R.,  2013, \mn@doi [ApJ]
  {10.1088/0004-637X/768/2/175}, 768, 175

\bibitem[\protect\citeauthoryear{Taylor, Gugliucci, Fabian, Sanders, Gentile
  \& Allen}{Taylor et~al.}{2006}]{Taylor2006}
Taylor G.~B.,  Gugliucci N.~E.,  Fabian A.~C.,  Sanders J.~S.,  Gentile G.,
  Allen S.~W.,  2006, \mn@doi [MNRAS] {10.1111/j.1365-2966.2006.10244.x}, 368,
  1500

\bibitem[\protect\citeauthoryear{Thierbach, Klein  \& Wielebinski}{Thierbach
  et~al.}{2003}]{Thierbach2003}
Thierbach M.,  Klein U.,   Wielebinski R.,  2003, \mn@doi [Astronomy \&
  Astrophysics] {10.1051/0004-6361}, 61, 53

\bibitem[\protect\citeauthoryear{Voigt \& Fabian}{Voigt \&
  Fabian}{2004}]{Voigt2004}
Voigt L.~M.,  Fabian A.~C.,  2004, \mn@doi [MNRAS]
  {10.1111/j.1365-2966.2004.07285.x}, 347, 1130

\bibitem[\protect\citeauthoryear{Vorst}{Vorst}{1992}]{Vorst1992}
Vorst H. V.~D.,  1992, SIAM Journal on Scientific and Statistical Computing,
  13, 631

\bibitem[\protect\citeauthoryear{Yang \& Reynolds}{Yang \&
  Reynolds}{2015}]{Yang2015}
Yang H. Y.~K.,  Reynolds C.~S.,  2015, preprint (\mn@eprint {arXiv}
  {1512.05796})

\bibitem[\protect\citeauthoryear{Zakamska \& Narayan}{Zakamska \&
  Narayan}{2003}]{Zakamska2003}
Zakamska N.~L.,  Narayan R.,  2003, ApJ, 582, 162

\bibitem[\protect\citeauthoryear{ZuHone, Markevitch, Ruszkowski  \& Lee}{ZuHone
  et~al.}{2013}]{ZuHone2013}
ZuHone J.~a.,  Markevitch M.,  Ruszkowski M.,   Lee D.,  2013, \mn@doi [ApJ]
  {10.1088/0004-637X/762/2/69}, 762, 69

\bibitem[\protect\citeauthoryear{ZuHone, Kunz, Markevitch, Stone  \&
  Biffi}{ZuHone et~al.}{2015}]{ZuHone2015}
ZuHone J.~a.,  Kunz M.~W.,  Markevitch M.,  Stone J.~M.,   Biffi V.,  2015,
  \mn@doi [ApJ] {10.1088/0004-637X/798/2/90}, 798, 90

\makeatother
\end{thebibliography}

\appendix
\section{Solving the Taylor approximation for the second order term}
	\label{appendixsecondorder}
	
	We show the solution of the modified Eq. (\ref{EQqnumericstaylor}) solved for the second order term, which we need to compute the mixed second derivatives in the conduction equation.
	At first, the kernel derivative is expressed as
	\begin{equation}
	\frac{\partial \Wij}{\partial \lp \xi \rp_\delta} = - \Wij' \frac{\lp \xij \rp_\delta}{\abs{\xij}}.
	\end{equation}
	
	Looking at the first order error term of the modified Eq. (\ref{EQqnumericstaylor}), we see, that
	\begin{equation}
	\int d^3 \xj ~ \frac{\lp \xij \rp_\alpha \lp \xij \rp_\gamma}{\abs{\xij}^2} \cdot \lp - \frac{\lp \xij \rp_\delta}{\abs{\xij}} \Wij' \rp = 0,
	\end{equation}
	since for all possibilities of $\alpha$, $\gamma$ and $\delta$ there is always at least one component, where the integral vanishes because of an antisymmetric integrand. All indices range from 1 to 3, so there is always one component with an odd amount of $\xij$. The denominator and $\Wij'$ are even with respect to $\xj$, so the integral vanishes.
	
	The next step is to calculate the integrals of the second order error term with a substitute as
	\begin{equation}
	T_{\alpha \beta \gamma \delta} = \int d^3 \xj ~ \frac{\lp \xij \rp_\alpha \lp \xij \rp_\beta \lp \xij \rp_\gamma \lp \xij \rp_\delta}{\abs{\xij}^3} \cdot \Wij'.
	\label{EQsecondordererrorterm}
	\end{equation}
	
	We distinguish between the following three cases, which we address one after another:
	\begin{enumerate}
		\item At least three indices are unequal
		\item All indices are equal
		\item The indices form two pairs, e.g. $\alpha = \beta$ and $\gamma = \delta$
	\end{enumerate}
	
	\subsection*{1. At least three indices are unequal}
	
		If at least three of the four indices are unequal, then there is at least one integration where the integrand contains only a single $\xij$ component. Since the denominator and $\Wij'$ are even functions with respect to $\xj$, the integrand for this component is in total an odd function which vanishes when integrating over the whole (symmetric) domain. Therefore, the integral is zero.
		
	\subsection*{2. All indices are equal}
	
		If all indices are equal, we start the calculations with substituting the integration variable $\xj \rightarrow \xij$ without further implications on the integration. Then Eq. (\ref{EQsecondordererrorterm}) simplifies to
		\begin{equation}
			T_\alpha = \int d^3 \xij ~ \frac{\lp \xij \rp_\alpha^4}{\abs{\xij}^3} \cdot \Wij',
		\end{equation}
		where we used the short hand notation $T_\alpha := T_{\alpha \alpha \alpha \alpha}$.
		
		Since $\Wij'$ is only dependent on $\abs{\xij}$ we choose spherical coordinates for $\xij$. We can arbitrarily choose the rotation of our coordinate system. For simplicity we let $\lp \xij \rp_\alpha$ be along the $z$-axis of the coordinate system. This results in
		\begin{equation}
			T_{\alpha} = \int dr \int d \phi \int d \theta ~ r^2 \sin\theta \cdot \lp r \cos \theta \rp^4 \cdot \frac{W'(r)}{r^3}.
		\end{equation}
		We can easily perform the $\phi$ and $\theta$ integrations and obtain
		\begin{equation}
			T_{\alpha} = \frac{4 \pi}{5} \int dr ~ r^3 W'(r).
		\end{equation}
		Next, we perform a partial integration, where the boundary term vanishes, since the kernel is monotonically decreasing towards zero. It remains:
		\begin{equation}
			T_{\alpha} = - \frac{12 \pi}{5} \int dr ~ r^2 W(r) = - \frac{3}{5},
		\end{equation}
		because of the kernel normalisation condition.
		
	\subsection*{3. The indices form two pairs}
	
		This last case can be calculated pretty similar, except that we have to chose two indices, which have to be unequal. We choose $\alpha = 1$ and $\beta = 3$. Again, written in spherical coordinates we get
		\begin{equation}\begin{array}{ll}
			T_{\alpha \beta} = \int dr \int d \phi \int d \theta ~& r^2 \sin\theta \cdot \lp r \sin\theta \cos\phi \rp^2 \cdot\\
			& \lp r \cos\theta \rp^2 \cdot \frac{W'(r)}{r^3}.
		\end{array}\end{equation}
		Using the results from the $r$-integration before we calculate \mbox{$T_{\alpha \beta} = - \frac{1}{5}$} and the total result is:
		
	\renewcommand{\arraystretch}{1.3}
	\begin{equation}
		T_{\alpha \beta \gamma \delta} =
		\left\{\begin{array}{ll}
		-\frac{3}{5} &\mbox{if } \alpha = \beta = \gamma = \delta\\
		-\frac{1}{5} &\mbox{if } \alpha = \beta \ne \gamma = \delta\\
		-\frac{1}{5} &\mbox{if } \alpha = \gamma \ne \beta = \delta\\
		-\frac{1}{5} &\mbox{if } \alpha = \delta \ne \beta = \gamma\\
		0 &\mbox{else}
		\end{array}\right.
		\label{EQsecondordererrortermresult}
	\end{equation}
	\renewcommand{\arraystretch}{1.0}
	So basically $T$ is only non zero, if we have two pairs of indices.
	
	\subsection*{All cases combined}
	
	Plugging everything back into the modified Eq. (\ref{EQqnumericstaylor}) gives:
	\begin{equation}\begin{array}{ll}
		I_{\gamma \delta} & := 2 \int d^3 \xj ~ \frac{Q \lp \xj \rp - Q \lp \xi \rp}{\abs{\xij}^2} \lp \xij \rp_\gamma \frac{\partial \Wij}{\partial \lp \xi \rp_\delta}\\
		&= - \sum\limits_{\alpha \beta} T_{\alpha \beta \gamma \delta} \at{\frac{\partial^2 Q}{\partial x_\alpha \partial x_\beta}}{\xi}.
	\end{array}\end{equation}
	To infer a general behaviour we take a look at an example with $\gamma = \delta = 0$
	\begin{equation}
		I_{0 0} = \frac{3}{5} \at{\frac{\partial^2 Q}{\partial^2 x_0}}{\xi} + \frac{1}{5} \at{\frac{\partial^2 Q}{\partial^2 x_1}}{\xi} + \frac{1}{5} \at{\frac{\partial^2 Q}{\partial^2 x_2}}{\xi}.
	\end{equation}
	Since we want to infer an approximation for second order mixed derivatives of $Q$, we have to linearly combine terms for different choices of $\gamma$ and $\delta$. It can be found that,
	\begin{equation}\begin{array}{ll}
		\frac{\partial^2 Q}{\partial x_0^2} &= 2 \cdot I_{0 0} - \frac{1}{2} \cdot I_{1 1} - \frac{1}{2} \cdot I_{2 2}\\
		&= 2 \cdot \frac{5}{4} \cdot I_{0 0} - \frac{1}{2} \cdot I_{0 0} - \frac{1}{2} \cdot I_{1 1} - \frac{1}{2} \cdot I_{2 2}
		\label{EQsecondordernonmixed}
	\end{array}\end{equation}
	and cyclic permutation for other second derivatives. A similar formula applies for mixed derivatives. Consider for example
	\begin{equation}
		I_{0 1} = \frac{1}{5} \at{\frac{\partial^2 Q}{\partial x_0 \partial x_1}}{\xi} + \frac{1}{5} \at{\frac{\partial^2 Q}{\partial x_1 \partial x_0}}{\xi}.
	\end{equation}
	It is better to keep both parts separated to explicitly indicate the symmetry for simplicity in the assembly process.
	
	From this example we get directly
	\begin{equation}
		\frac{\partial^2 Q}{\partial x_0 \partial x_1} = \frac{5}{4} \cdot I_{0 1} + \frac{5}{4} \cdot I_{1 0}
		\label{EQsecondordermixed}
	\end{equation}
	and cyclic permutations.
	
	Combining the equations (\ref{EQsecondordernonmixed}) and (\ref{EQsecondordermixed}) we see that all $I_{\gamma \delta}$ occur times a factor of 5/2 minus a trace term times 1/2. This finding is represented by the substition Eq. (\ref{EQtildevariabledef}) in our final result.
	
\section{Analytic solution for the temperature step problem}
	\label{appendixtemperaturestep}
	
	We show the derivation of the analytic solution of the conduction equation for the temperature step test (section \ref{temperaturestep}).
	We start with the Fourier transformation of the specific internal energy in both directions:
	
	\begin{equation}
		u(t, x) = \infint u_k(t) e^{ikx} ~ 	\frac{dk}{2\pi} \label{EQfourierbacktrafo}
	\end{equation}
	and
	\begin{equation}
		u_k(t) = \infint u(t, x) e^{-ikx} ~ dx .
	\end{equation}
	The conduction equation expressed in Fourier space is
	\begin{equation}
		\frac{du_k(t)}{dt} = -\alpha k^2 u_k(t)
	\end{equation}
	with the simple solution
	\begin{equation}
		u_k(t) = u_{k0} e^{-\alpha k^2 t}.
	\end{equation}
	Using $u(t=0, x) = u_0(x)$ we express the unknown coefficient in terms of the initial condition in real space
	\begin{equation}
		u_{k0} = \infint u_0(x') e^{-ikx'} dx'.
	\end{equation}
	We insert this result into the reverse Fourier transformation (Eq. (\ref{EQfourierbacktrafo})) and obtain
	\begin{equation}
		u(t,x) = \infint dx' \infint \frac{dk}{2\pi} \; u_0(x') e^{-\alpha k^2 t} e^ {ik(x-x')}.
	\end{equation}
	At first, we perform the integration over $dk$. For this we rewrite the exponentials completing the square to bring them into Gaussian form, which is a simple integration and get
	\begin{equation}
		u(t,x) = \frac{1}{2\sqrt{\pi \alpha t}} \infint dx' \; u_0(x') \exp \lp - \frac{(x-x')^2}{4 \alpha t} \rp . \label{EQlastgeneralsolutionstep}
	\end{equation}
	At this point we need to use the specific initial conditions of our problem. For the temperature step they are defined as
	\renewcommand{\arraystretch}{1.3}
	\begin{equation}
		u_0(x') =  
		\left\{\begin{array}{ll}
		u_0 - \frac{\Delta u}{2} & \mbox{for } x' < x_m\\
		u_0 + \frac{\Delta u}{2} & \mbox{for } x' > x_m,
		\end{array}\right.
		\label{EQstepics}
	\end{equation}
	\renewcommand{\arraystretch}{1.0}
	with $x_m$ being the position of the temperature step.
	
	Inserting this into Eq. (\ref{EQlastgeneralsolutionstep}) we get two integrals to perform:
	\begin{equation}\begin{array}{ll}
		u(t,x) = \frac{1}{2\sqrt{\alpha \pi t}} & \lb \int\limits_{-\infty}^{x_m}dx' \; \lp u_0 - \frac{\Delta u}{2} \rp e^{-y^2} + \right.\\
		&\hspace{6pt} \left. \int\limits_{x_m}^{\infty}dx' \; \lp u_0 + \frac{\Delta u}{2} \rp e^{-y^2} \rb , \label{EQpluggedintemperaturestep}
	\end{array}\end{equation}
	where we substituted ~~$y \equiv y(x') = \sqrt{\frac{(x-x')^2}{4 \alpha t}}$~.
	
	We split this expression into two parts: One which is multiplied by $u_0$ and one which is multiplied by $\Delta u / 2$. The $u_0$ term can be simply integrated, since it is again only a gaussian integral, which results in $2\sqrt{\alpha \pi t} \cdot u_0$. For a little consistence check consider $\Delta u = 0$ then we get $u(t, x) = u_0$, which is what we would expect for a isothermal region without any other effects than thermal conduction.
	
	To integrate the second term, with $\Delta u$, one has to rewrite the integrals to resemble the definition of error functions. The final result reads
	\begin{equation}
		u(t,x) = u_0 + \frac{\Delta u}{2} \cdot \erf \lp \frac{x-x_m}{2\sqrt{\alpha t}} \rp .
	\end{equation}
	
\section{Temperature step with vacuum boundaries}
	\label{appendixtemperaturestepbounds}
	
	\nr{
	In section \ref{timeevolutiontest} we require a similar derivation as shown in appendix \ref{appendixtemperaturestep}, however, we need to change the boundaries of the initial conditions given by Eq. \ref{EQstepics} to
	\renewcommand{\arraystretch}{1.3}
	\begin{equation}
	u_0(x') =  
	\left\{\begin{array}{ll}
	u_0 - \frac{\Delta u}{2} & \mbox{for } x_0 < x' < x_m\\
	u_0 + \frac{\Delta u}{2} & \mbox{for } x_m < x' < x_1\\
	0 & \mbox{otherwise.}
	\end{array}\right.
	\end{equation}
	\renewcommand{\arraystretch}{1.0}
	This leads to finite integral boundaries in the $x$-integration which be written again in terms of the error function
	\begin{equation}\begin{array}{ll}
	u(t,x) &= u_0 \cdot \left[ \erf \lp y \lp t, x_1 \rp \rp + \erf \lp y \lp t, x_0 \rp \rp \right] \nonumber\\
	&+ \frac{\Delta u}{4} \cdot \left[ - \erf \lp y \lp t, x_1 \rp \rp + 2 \cdot \erf \lp y \lp t, x_m \rp \rp - \erf \lp y \lp t, x_2 \rp \rp \right]
	\end{array}\end{equation},
	which behaves perfectly fine for $x_0 \rightarrow - \inf$, $x_1 \rightarrow \inf$. However, if we let $t \rightarrow \inf$ then $y \rightarrow 0$ and therefore $u \rightarrow 0 \ne u_0$. Mathematically the difference lies in two integrals of infinite integration length with integrand zero. Physically this can be understood as heat dissipating into the boundaries given by $x_0$ and $x_1$. Therefore, it is not possible in a straight forward way to extract an analytic solution for these modified, finite initial temperature step.
	}

\label{lastpage}

\end{document}